\documentclass[preprint,12pt,authoryear]{elsarticle}

\usepackage{amsmath,amssymb}
\usepackage{tikz}
\usetikzlibrary{shapes.geometric, arrows,calc,trees,positioning,arrows,chains, ,decorations.pathmorphing,shapes, matrix,shapes.symbols,matrix}
\usepackage{tabularx} 
\allowdisplaybreaks
\usepackage{pgfplots}
\pgfplotsset{compat=1.18}
\usepackage{enumitem}
\usepackage{multirow}
\usepackage{mathtools}
\usepackage{empheq}
\usepackage{array}
\usepackage{rotating}
\usepackage{adjustbox}
\usepackage[raggedrightboxes]{ragged2e}
\usepackage{float}
\usepackage{natbib}
\usepackage[colorlinks=true, allcolors=blue]{hyperref}
\usepackage{caption}  
\usepackage{colortbl}
\definecolor{highlightcolor}{RGB}{255, 255, 0}
\usepackage{booktabs}%
\usepackage{graphicx}
\usepackage{url}
\usepackage{breakurl}

\journal{Energy Reports}

\begin{document}

\begin{frontmatter}

\title{Analysis of Short-Run and Long-Run Marginal Costs of Generation in the Power Market} 

\author[label1]{Shamim Homaei}
\author[label2]{Simon Roussanaly}
\author[label1]{Asgeir Tomasgard}
\affiliation[label1]{organization={Department of Industrial Economics and Technology Management, Norwegian University of Science and Technology},
             city={Trondheim},
             country={Norway}}

 \affiliation[label2]{organization={SINTEF Energy Research},
             city={Trondheim},
             country={Norway}}

\begin{abstract}
In power markets, understanding the cost dynamics of electricity generation is crucial. The complexity of price formation in the power system arises from its diverse attributes, such as various generator types, each characterized by its specific fixed and variable costs as well as different lifetimes. In this paper, we adopt an approach that investigates both long-run marginal cost (LRMC)  and short-run marginal cost (SRMC) in a perfect competition market. According to economic theory, marginal pricing serves as an effective method for determining the generation cost of electricity. This paper presents a capacity expansion model designed to evaluate the marginal cost of electricity generation, encompassing both long-term and short-term perspectives. Following a parametric analysis and the calculation of LRMCs, this study investigates the allocation of investment costs across various time periods and how these costs factor into the LRMC to ensure cost recovery. Additionally, an exploration of SRMCs reveals the conditions under which LRMCs and SRMCs converge or diverge. We observe that when there is a disparity between LRMC and SRMC, setting electricity generation prices equal to SRMCs does not ensure the complete recovery of investment and operational costs. This phenomenon holds implications for market reliability and challenges the pricing strategies that rely solely on SRMCs. Furthermore,  our investigation highlighted the significance of addressing degeneracy in the power market modeling. Primal degeneracy in the SRMC model can result in multiple values for the dual variable representing SRMC. This multiplicity of values creates ambiguity regarding the precise SRMC value, making it challenging to ascertain the correct estimation. As a result, resolving degeneracy will ensure the reliability of the SRMC value, consequently enhancing the robustness and credibility of our analysis.
\end{abstract}

\begin{highlights}
\item Presenting a method for calculating long-run marginal costs of electricity generation.
\item Exploring the relationship between long-run and short-run marginal costs of electricity generation.
\item Assessing the feasibility of cost recovery under long-run and short-run marginal cost pricing methods.
\item Investigating the role of degeneracy in calculating short-run marginal costs of electricity generation.
\end{highlights}

\begin{keyword}
Power market \sep Marginal pricing \sep Cost allocation \sep Cost recovery \sep Capacity expansion
\end{keyword}

\end{frontmatter}



\section{Introduction}

In power markets, a fundamental question arises: what constitutes the cost of electricity generation, and which generator technologies and capacities should be installed to optimize this cost? However, this inquiry is intricate due to several factors. Diverse types of generators each bear their distinct variable and fixed costs, along with unique lifetimes, posing challenges in determining the cost of electricity generation \citep{MALIK20061703}.

According to economic theory, the optimal pricing strategy for various commodities, including electricity, aligns with the marginal cost. In the power market, where the primary objective is to maximize social welfare,  it is the marginal generation cost that determines the price of electricity under the assumption of perfect competition \citep{berrie1992,MALIK20061703,pikk2013dangers}. Marginal cost is the cost incurred by the electricity system when there is a rise in electricity demand within a specific time and region. The concept of marginal cost can be examined from both long-run and short-run perspectives.

When computing the long-run marginal cost (LRMC), all generation factors are treated as variables, capable of adjustment in response to additional demand. The LRMC reflects the future impact of this additional demand on the overall power system cost trajectory, including both investment and operational expenses. In contrast, the short-run marginal cost (SRMC) calculation assumes a fixed invested capacity and focuses on the cost of supplying an additional unit using the existing capacity \citep{MALIK20061703,GOVAERTS2023101537}. 

The significance of LRMC lies in its ability to ensure the recovery of generation costs \citep{conejo2013}. However, the use of SRMC does not always guarantee cost recovery and may result in missing money problems \citep{wogrin2022impact}. There are instances where LRMC and SRMC align, but this convergence depends on specific conditions. It is important to understand the relationship between LRMC and SRMC and identify the circumstances under which they coincide. This understanding can aid in determining a more appropriate method for pricing electricity and how investment costs are allocated among different periods to secure cost recovery. Each generator type has its own operational lifetime that spans multiple periods. Thus, understanding how investment costs are spread out during the generator's active periods and how such allocation impacts LRMCs in these periods is essential. 

In the related literature, numerous methods have been proposed for estimating marginal cost. These methods can be categorized into two primary approaches: Marginal Incremental Cost (MIC) and Average Incremental Cost (AIC). These approaches have been investigated in works by \citet{marsden2004estimation,ekwue2014assessment,tooth2014measuring}. The MIC defines marginal cost as the change in total costs, encompassing both capacity and operational costs, associated with a unit change in demand within a single period. In contrast, the AIC takes into account incremental shifts in demand spanning multiple planning periods. Based on the AIC, marginal cost is computed as the average cost incurred to meet these fluctuations in demand. Several research studies, including works by \citet{porat1997long,lima2002distribution,de2003solving,MALIK20061703}, and \citet{GOVAERTS2023101537}, have utilized the AIC method for estimating marginal costs. In our paper, we employ the MIC approach to determine the LRMC of electricity generation. This choice aligns with one of our primary objectives, which is to understand the dynamics of the formation of electricity prices. The MIC is particularly effective in isolating the influence of individual factors within complex systems. Thus, using this approach is beneficial as it allows us to investigate the impact of a singular factor, specifically the increase in demand, on the dynamics of electricity pricing. Furthermore, the MIC method is considered a more reliable approach for estimating LRMC in comparison to the AIC method \citep{tooth2014measuring}.

The shadow pricing is a commonly employed method to estimate marginal costs, as evidenced by foundational works such as \citet{mathiesen1977marginal} and \citet{munasinghe1982electricity}. This method is classified within the MIC framework. By employing shadow pricing, we calculate the dual values associated with various constraints. Particularly, in capacity expansion models,  the dual value corresponding to the flow-balance constraint allows us to determine the LRMC of electricity generation. This dual value concept allows us to understand how total costs change when there is a shift in demand within a specific period. The shadow pricing method has been widely employed in numerous studies, including those conducted by researchers such as \citet{ring1995dispatch,remme2014dual,munoz2018economic,korpaas2020optimality} and \citet{wogrin2022impact}. Despite the advantages associated with shadow pricing, its application may give rise to a specific challenge known as degeneracy, which is examined in this paper.

In conditions of perfect competition, market prices are determined by the marginal cost of production \citep{pikk2013dangers}. In a study by \citet{conejo2013}, the role of marginal cost pricing in recovering investment costs was examined. Conejo employed a simplified model of a perfectly competitive power market, with the primary goal being the maximization of social welfare, and assuming inelastic demand. The study demonstrated that the profits generated by infra-marginal technologies are sufficient to cover their investment costs. In another study, \citet{korpaas2020optimality} investigated the issue of cost recovery in the context of a perfectly competitive electricity market. They employed the load duration curve approach with the objective of minimizing the system cost, focusing on scenarios with constant demand levels. The findings demonstrated that the SRMC alone is capable of recovering investment expenditures in the electricity market. \citet{rodriguez2010regulatory} employed a model aimed at maximizing social benefits to assess the effectiveness of SRMC in providing efficient signals for optimal long-term investments. This analysis was conducted under the assumptions of a perfect market, continuous investment, and convex cost functions. The findings indicate that SRMC can recover investment costs in cases where the marginal technology is not operating at its capacity limit and there is no loadshed. 

\citet{wogrin2022impact} investigated the potential of SRMC in recovering investment expenditures. They employed a simplified capacity expansion model comprising one generator and one time period. SRMC was derived using dual calculations for the flow balance constraint while maintaining fixed capacity variables. Their findings suggest that SRMC is inadequate for covering investment costs, a limitation attributed to degeneracy within the primal model. Although degeneracy is a well-recognized phenomenon in linear programming,  its significance in practical applications, such as power systems,  is frequently underestimated \citep{gamrath2020exploratory}. Given the managerial importance of addressing degeneracy issues in power market models, our study examines this aspect. 

In this paper, we develop a methodology for calculating the LRMC of electricity generation. Our approach begins with a capacity expansion model, followed by deriving the dual model and applying complementary slackness conditions to determine the LRMCs. To capture the full range of potential LRMC values, we identify all relevant instance groups and explore the resulting LRMC combinations. We then analyze the formation of LRMC, how investment costs are distributed across different periods, and whether cost recovery is achievable under LRMC-based pricing. Additionally, we calculate the SRMC and investigate the conditions under which LRMC and SRMC converge or diverge. Our study also examines the feasibility of SRMC in covering generation expenses, the impact of degeneracy on SRMC, and strategies for resolving degeneracy. This methodology not only deepens our understanding of LRMC dynamics but also enables the evaluation of cost recovery feasibility and the optimization of resource allocation within an LRMC-based pricing framework, while exploring its relationship with SRMC.

The subsequent sections of the paper are structured as follows: Section \ref{sec:LRMC estimation methodology} outlines the employed methodology for estimating LRMC, followed by an analysis of the estimated LRMC values in Section \ref{sec:Analysis of LRMC combinations}. Section \ref{sec:SRMC estimation methodology} details the methodology for estimating SRMC values. Finally, Section \ref{sec:Concluding remarks} concludes the paper and discusses future avenues for further exploration.

\section{LRMC estimation methodology} \label{sec:LRMC estimation methodology}

In this section, we present a methodology for calculating the LRMC in the context of a capacity expansion model for the electricity sector. This methodology consists of the following steps:
\begin{enumerate} [label=Step \arabic*. ,wide]
    \item Presenting the capacity expansion model: In the initial step, we introduce the stylized capacity expansion model for the power market. This model forms the foundation of our analysis, providing the framework to optimize capacity decisions while determining the optimal electricity mix (see Subsection \ref{sec:Primal model for long-run marginal cost}).
    \item Deriving the dual model: Our primary aim is to determine the LRMC, which is linked to the dual value of the flow-balance constraint in the capacity expansion model. To achieve this, we construct the dual model of the capacity expansion model (see Subsection \ref{sec:Dual model for long-run marginal cost}).
    \item Establishing primal-dual relationships: It is imperative to establish a linkage between the primal and dual models. This connection is achieved through the formulation of complementary slackness conditions, which provide valuable insights for extracting the LRMC values (detailed in Subsection \ref{sec:Complementary slackness}).
    \item Identifying diverse LRMC instance groups: In this step, we aim to identify all potential LRMC instance groups based on specific conditions. To do this, we examine four critical factors: peak demand period, marginal generation option in off-peak demand period, loadshed cost, and invested capacity limit (refer to Subsection \ref{sec:Identification of LRMC cases}). Accordingly, we can find a specific LRMC value for each instance group.

    \item Categorizing LRMC combinations: After determining the LRMC values for each instance group, it was observed that certain LRMC values recurred across multiple cases, leading to their categorization into 7 unique combinations. This step focuses on the exploration of these distinct LRMC combinations, aiming to uncover the specific conditions that give rise to each combination and to provide insights into the factors shaping the formation of each LRMC combination (refer to Section \ref{sec:LRMC profiles}).
\end{enumerate}

This structured methodology not only deepens our understanding of LRMC dynamics but also assists in evaluating the feasibility of cost recovery within the LRMC pricing framework. It also provides valuable insights into the allocation of investment costs across different periods.

\subsection{Step 1: Primal model for LRMC}\label{sec:Primal model for long-run marginal cost}

The foundation of the proposed methodology is a capacity expansion model for the power sector. The objective of the model is minimizing the overall cost of the system while determining the most efficient investment strategy for power generation and operational generation decisions. This model considers two types of generators and two time periods. Moreover, the model assumes a perfectly competitive market, continuous investment decisions, and inelastic demands. The capacity expansion model will be referred to as the 'primal model for LRMC' hereafter. A description of its sets, parameters, and variables can be found in Table \ref{tab:Description_of_sets_parameters_and_variables_in_primal_model_for_LRMC}. Additionally, we assume that the investment and operational costs associated with fossil-based are greater than those of renewable generators. 
\begin{table}[h]
\centering
\caption{Description of sets, parameters, and variables in primal model for LRMC}
\label{tab:Description_of_sets_parameters_and_variables_in_primal_model_for_LRMC}
\begin{tabular}{ >{\raggedright}p{2.5cm} p{9cm} }
\hline
  Sets & \\
  \hline
  $g \in G$ & Generation technologies;  $G = \{f, r\}$ \newline ($f$: Fossil-based generator; $r$: Renewable generator) \\
  $t \in T$ & Time periods; $T = \{1, 2\}$ \\
  $u_g \in \{u_f, u_r\}$ & Lifetime of generator $g$; $(u_f = 2, u_r = 2)$ \\
  \hline
  Parameters &  \\
  \hline
  $CI_{g}$ & Investment cost of generator $g$ \\
  $CP_{g}$ & Operational cost of generator $g$ \\
  $CL$ & Loadshed cost \\
  $D_{t}$ & Demand in period $t$ \\
  $M_{g}$ & Maximum investable capacity of generator $g$ \\
  \hline
  Variables & \\
  \hline
  $I_{gt}$ & Invested capacity of generator $g$ in period $t$ \\
  $P_{gt}$ & Electricity generation by generator $g$ in period $t$ \\
  $L_t$ & Loadshed in period $t$ \\
  \hline
\end{tabular}
\end{table}

The primal model for the LRMC model is as follows:
\begin{align}
   \min z = \sum_{g,t} CI_g \cdot I_{gt} + \sum_{g,t} CP_g \cdot P_{gt} + \sum_t CL \cdot L_t \qquad\qquad \label{eq:OF}
\end{align}
\begin{alignat}{2}
    s.t.  &\quad \sum_g P_{gt} + L_t = D_t &&\quad \forall t \in T \label{eq:flow-balance} \\
          &\quad P_{gt} \leq \sum_{t'=t''}^{t} I_{gt'} \quad &&\quad t''=\max\{1,t-u_g+1\};  \forall g \in G, t \in T \label{eq:capacity} \\
          &\quad I_{gt} \leq M_g &&\quad \forall g \in G, t \in T \label{eq:capacity max} \\
          &\quad P_{gt} \geq 0, I_{gt} \geq 0, L_{t} \geq 0 &&\quad \forall g \in G, t \in T \label{eq:sign} 
\end{alignat}

The objective function, Equation \ref{eq:OF}, minimizes the total cost of the system over the planning horizon, including investment and operational costs of generators and the costs incurred due to loadshed. Constraint \ref{eq:flow-balance} represents the balance in electricity flow for each period. Constraint \ref{eq:capacity} ensures that any generator's electricity generation during a specific period does not surpass its total installed capacity by that period. Furthermore, Constraint \ref{eq:capacity max} imposes a limitation on the capacity that can be invested for each generator in each period.

This model has been formulated in its compact form. We transition the model to its standard form to facilitate the interpretation of dual variables in the next step. In addition, we have assigned a dual variable to each constraint, which will be explained in more detail in Subsection \ref{sec:Dual model for long-run marginal cost}. Finally, the expanded and standard form of the primal model for LRMC is as follows:
\begin{align}
    \min z &= CI_r \cdot I_{r1} + CI_r \cdot I_{r2} + CI_f \cdot I_{f1} + CI_f \cdot I_{f2} + CP_r \cdot P_{r1} + CP_r \cdot P_{r2} \label{eq:primal OF} \notag\\ 
    &\quad +  CP_f \cdot P_{f1} + CP_f \cdot P_{f2} + CL \cdot L_1 + CL \cdot L_2 && 
\end{align}
\begin{alignat*}{2}
    s.t.  &\quad P_{r1} + P_{f1} + L_1 = D_1 && \quad (\lambda_1 \in (-\infty, +\infty)) \\
          &\quad P_{r2} + P_{f2} + L_2 = D_2 && \quad (\lambda_2 \in (-\infty, +\infty)) \\
          &\quad -P_{r1} + I_{r1} \geq 0 && \quad (\beta_{r1} \geq 0)\\
          &\quad -P_{r2} + I_{r1} + I_{r2} \geq 0 && \quad (\beta_{r2} \geq 0) \\
          &\quad -P_{f1} + I_{f1} \geq 0 && \quad (\beta_{f1} \geq 0)     \\  
          &\quad -P_{f2} + I_{f1} + I_{f2} \geq 0 && \quad (\beta_{f2} \geq 0) \\
          &\quad -I_{r1} \geq -M_r && \quad (\gamma_{r1} \geq 0)\\
          &\quad -I_{r2} \geq -M_r && \quad (\gamma_{r2} \geq 0)\\
          &\quad -I_{f1} \geq -M_f && \quad (\gamma_{f1} \geq 0)\\
          &\quad -I_{f2} \geq -M_f && \quad (\gamma_{f2} \geq 0)\\
          &\quad P \geq 0; I \geq 0; L \geq 0 
\end{alignat*}

\subsection{Step 2: Dual model for LRMC}\label{sec:Dual model for long-run marginal cost}

The dual problem associated to the primal model for LRMC (Equation \ref{eq:primal OF}) is outlined in Equation \ref{eq:dual OF}. This particular model is referred to as the 'dual model for LRMC.'
\begin{align}
    \max y &= D_1 \cdot \lambda_{1} + D_2 \cdot \lambda_{2} - M_r \cdot \gamma_{r1} - M_r \cdot \gamma_{r2} - M_f \cdot \gamma_{f1} - M_f \cdot \gamma_{f2} && \label{eq:dual OF}
\end{align}
\begin{alignat*}{2}
    s.t.  &\quad \lambda_{1} - \beta_{r1} \leq CP_r && \quad (P_{r1} \geq 0) \\
          &\quad \lambda_{2} - \beta_{r2} \leq CP_r && \quad (P_{r2} \geq 0) \\
          &\quad \lambda_{1} - \beta_{f1} \leq CP_f && \quad (P_{f1} \geq 0) \\
          &\quad \lambda_{2} - \beta_{f2} \leq CP_f && \quad (P_{f2} \geq 0) \\
          &\quad \beta_{r1} + \beta_{r2} - \gamma_{r1} \leq CI_r && \quad (I_{r1} \geq 0)     \\  
          &\quad \beta_{r2} - \gamma_{r2} \leq CI_r && \quad (I_{r2} \geq 0) \\
          &\quad \beta_{f1} + \beta_{f2} - \gamma_{f1} \leq CI_f && \quad (I_{f1} \geq 0)     \\  
          &\quad \beta_{f2} - \gamma_{f2} \leq CI_f && \quad (I_{f2} \geq 0) \\
          &\quad \lambda_1 \leq CL && \quad (L_1 \geq 0)\\
          &\quad \lambda_2 \leq CL && \quad (L_2 \geq 0)\\
          &\quad \beta \geq 0; \gamma \geq 0; \lambda \in (-\infty, +\infty)
\end{alignat*}
Each constraint in Equation \ref{eq:primal OF} has been assigned a dual variable. $\lambda_t$ is the dual variable of the flow-balance constraint for each time period $t$. $\lambda_t$ serves as an indicator of the LRMC of electricity generation during time period $t$. It shows the cost of generating an additional unit of electricity, accounting for both operational and investment costs. Furthermore, $\beta_{gt}$ is assigned as the dual value of the generation capacity constraint. This dual variable represents the change in cost resulting from a reduction of one unit in the installed capacity during a specific period. In essence, $\beta$, referred to as capacity value, represents the economic benefit of excess capacity in meeting demand. 

Additionally, $\gamma$ is attributed to the dual value of the maximum invested capacity constraint, and it can be interpreted as opportunity cost. This dual value represents the benefit lost if the last unit of investable capacity was not available, hence requiring the adoption of a costlier alternative for additional generation.

\subsection{Step 3: Complementary slackness conditions}\label{sec:Complementary slackness}
So far, primal and dual models have been presented. This section aims to find a method of calculating LRMCs. The LRMC is equal to the dual of the flow-balance constraint. However, as both the primal and dual models are in parametric form, calculating the LRMC value directly is not possible. To address this, we aim to establish a relationship between the primal and dual models. Such a relationship can be demonstrated by complementary slackness. It states that in either the primal or dual model, the slack of a constraint and the corresponding dual variable are complementary, such that their product is always zero. In other words, if a constraint is not binding, the corresponding dual variable must be zero, and vice versa. For more details, refer to the study by \citet{bertsimas1997introduction}. Accordingly, the following are the complementary slackness conditions for the primal and dual models for LRMC:

\begin{alignat}{2}
Dual   &\quad P_{r1} \cdot (\lambda_{1} - \beta_{r1} - CP_r) = 0 \notag \\ 
       &\quad P_{r2} \cdot (\lambda_{2} - \beta_{r2} - CP_r) = 0 \notag \\
       &\quad P_{f1} \cdot (\lambda_{1} - \beta_{f1} - CP_f) = 0 \notag  \\
       &\quad P_{f2} \cdot (\lambda_{2} - \beta_{f2} - CP_f) = 0 \notag \\
       &\quad I_{r1} \cdot (\beta_{r1} + \beta_{r2} - \gamma_{r1} - CI_r) = 0 \notag \\  
       &\quad I_{r2} \cdot (\beta_{r2} - \gamma_{r2} - CI_r) = 0 \notag \\
       &\quad I_{f1} \cdot (\beta_{f1} + \beta_{f2} - \gamma_{f1} - CI_f) = 0 \notag \\  
       &\quad I_{f2} \cdot (\beta_{f2} - \gamma_{f2} - CI_f) = 0 \notag \\
       &\quad L_1 \cdot (\lambda_1 - CL) = 0 \notag \\
       &\quad L_2 \cdot (\lambda_2 - CL) = 0 \notag \\
Primal &\quad \lambda_1 \cdot (P_{r1} + P_{f1} + L_1 - D_1) = 0 \\
       &\quad \lambda_2 \cdot (P_{r2} + P_{f2} + L_2 - D_2) = 0 \notag\\
       &\quad \beta_{r1} \cdot (-P_{r1} + I_{r1}) = 0 \notag \\
       &\quad \beta_{r2} \cdot (-P_{r2} + I_{r1} + I_{r2}) = 0 \notag \\
       &\quad \beta_{f1} \cdot (-P_{f1} + I_{f1}) = 0 \notag \\  
       &\quad \beta_{f2} \cdot (-P_{f2} + I_{f1} + I_{f2}) = 0 \notag \\
       &\quad \gamma_{r1} \cdot (-I_{r1} + M_r) = 0 \notag \\
       &\quad \gamma_{r2} \cdot (-I_{r2} + M_r) = 0 \notag \\
       &\quad \gamma_{f1} \cdot (-I_{f1} + M_f) = 0 \notag \\
       &\quad \gamma_{f2} \cdot (-I_{f2} + M_f) = 0 \notag
\end{alignat}

\subsection{Step 4: Identification of LRMC instance groups}\label{sec:Identification of LRMC cases}

In the previous steps, we expressed both the primal and dual models in a parametric format, which resulted in the presentation of complementary slackness conditions in a similar manner. We categorize potential LRMC solutions based on four identification factors, aiming to cover all possible combinations.

Before exploring the identification factors, it is imperative to clarify the terminology used in this paper. We distinguish between 'generation technology' and 'generation option'. Generation technology pertains to the types of generators employed in electricity generation, specifically, renewable and fossil-based technologies. On the other hand, the generation option denotes specific configurations defined by the generation technology and its capacity structure. Two capacity structure are considered: 'shared capacity', invested in the first period and used in both periods, and 'non-shared capacity', invested in a period for exclusive use during that period. These distinctions give rise to four generation options: shared capacity of a renewable generator, non-shared capacity of a renewable generator, shared capacity of a fossil-based generator, and non-shared capacity of a fossil-based generator.

Moreover, it is important to distinguish between the terms 'average cost of electricity generation' and 'LRMC of electricity generation' in this paper. The average cost of electricity generation pertains to the real expenses incurred in the generation of electricity, encompassing both investment and operational costs as considered in this study. Notably, for the non-shared capacity, the average generation cost for one unit of electricity equals to the sum of investment cost and operational cost ($CI+CP$). This arises from the fact that a single unit of capacity is invested and utilized for generating a unit of electricity. In contrast, the average generation cost of electricity per unit using shared capacity is represented as $\frac{CI}{2}+CP$. The rationale behind this is the use of shared capacity, where one unit of shared capacity is invested and serves both the first and second periods, incurring operational costs for each period. On the other hand, the LRMC is determined through the marginal pricing method in this study. Hence, it is possible that the average generation cost and the LRMC associated with each period are not the same. 

The identification factors are as follows:

\begin{itemize}
  \item \textbf{Factor 1. Peak demand period}: It is crucial to determine which period experiences the highest demand, as this distinction leads to the formation of different instance groups. When the first period has the highest demand, the available generation options can include shared capacity and non-shared capacity of the first period. However, if the peak demand occurs in the second period, a broader spectrum of generation options becomes accessible to meet the demand. This includes shared capacity, non-shared capacity of the second period, and, non-shared capacity from the first period.

  \item \textbf{Factor 2. Marginal generation option in off-peak demand period}: Identifying the marginal generation option during the off-peak period is important. Off-peak demand can be efficiently met using the shared capacity of a generator type, granted that the capacity is sufficient, and consequently, load shedding does not occur. However, it is crucial to determine which type of generator, whether renewable or fossil-based, serves as the marginal option during the off-peak period. This differentiation impacts the calculation of LRMC and leads to the creation of distinct instance groups.  
  
  \item \textbf{Factor 3. Loadshed cost}: Loadshed cost plays a crucial role in determining whether it is more advantageous to generate electricity or resort to load shedding. This determination is made by comparing the average generation cost with the Loadshed cost. This comparative analysis reveals the array of potential generation options concerning load shedding costs, and it is the model itself that autonomously selects which generators to employ and determines their output. This identification factor is important because the LRMC varies depending on the presence or absence of load shedding and the availability of different generation options.
  
  \item \textbf{Factor 4. Invested capacity limit}: This factor evaluates whether the marginal generator during the peak demand period is operating at its maximum investable capacity. This distinction holds significant implications as, when the marginal generator is operating at full capacity and additional demand persists, it leads to capacity-constrained loadshed, consequently impacting the LRMC. This factor leads to different instance groups.  

\end{itemize}

\section{Step 5: Analysis of LRMC combinations} \label{sec:Analysis of LRMC combinations}

In the previous step, four identification factors were introduced. According to the four factors, we aim to identify all possible instance groups for a two-generator system. Initially, Factor 1 and Factor 2 are used to classify the instance groups into six larger groups, called 'clusters.' These clusters represent subsets of instance groups that share similar characteristics based on Factor 1 and Factor 2. Subsequently, the other two factors, Factor 3 and Factor 4, are employed to identify each single instance group within each cluster. This method leads to the discovery of a total of 41 instance groups across all clusters. All of the clusters, instance groups, and their related characteristics are presented in \hyperref[sec:supplementaryTables]{Appendix A}, Table \ref{tab:identification}. Since it is impractical to present all instance groups, we focus on a specific cluster to demonstrate how the identification factors contribute to the identification of instance groups. This includes detailing the characteristics of each instance group within the cluster and the related LRMC values for each instance group.

In cluster 1, regarding Factor 1, the second period is the peak demand period ($D_1 < D_2$), and regarding Factor 2, the demand in the off-peak period (first period) is below the maximum investable capacity of the renewable generator ($D_1 < M_r$). So, if generation is economically advantageous compared to loadshed, $D_1$ will be met by the renewable generator. Given the higher demand in the second period, the invested capacity from the first period can be used to meet the demand in both periods. Thus, this capacity functions as shared renewable capacity, fulfilling the entire demand in the first period and a portion of the demand in the second period.

In terms of Factor 3, the economic viability of a generation option depends on its cost relative to the loadshed cost ($CL$). If the average cost of a generation option's exceeds $CL$, it becomes economically unfeasible. For example, if $\frac{CI_r}{2}+CP_r < CL$, the total cost of generation using shared renewable capacity during both periods would be $CI_r+2CP_r$. However, if no investment is made in renewable, we will face loadshed in both periods with a cost of $2CL$. Since $CI_r+2CP_r < 2CL$, the more affordable option is generating with shared renewable capacity. This comparative analysis is summarized in Table \ref{tab:Generation option affordability based on loadshed cost}.

\begin{table}[htbp]
\centering
\caption{Generation option affordability based on loadshed cost}\label{tab:Generation option affordability based on loadshed cost}
\fontsize{9pt}{15pt}\selectfont  
\begin{tabular*}{0.85\textwidth}{@{\extracolsep\fill}lcccc}
\toprule
Generation type & \multicolumn{2}{c}{Renewable} & \multicolumn{2}{c}{Fossil-based} \\
\cmidrule(lr){2-3} \cmidrule(lr){4-5}
Capacity structure & Shared & Non-shared & Shared & Non-shared \\
\midrule
\hspace{56pt}$CL < \frac{CI_r}{2}+CP_r$                      &            &            &            &            \\
\hspace{0pt}$\frac{CI_r}{2}+CP_r < CL < \frac{CI_f}{2}+CP_f$ & \checkmark &            &            &            \\
\hspace{0pt}$\frac{CI_f}{2}+CP_f < CL < CI_r+CP_r$           & \checkmark &            & \checkmark &            \\
\hspace{1pt}$CI_r+CP_r < CL < CI_f+CP_f$                     & \checkmark & \checkmark & \checkmark &            \\
\hspace{1pt}$CI_f+CP_f < CL$                                 & \checkmark & \checkmark & \checkmark & \checkmark \\
\bottomrule
\end{tabular*}
\label{tab:CCSgeneration}
\end{table}

In cluster 1, depending on $CL$, shared renewable capacity can meet $D_1$, and a part of $D_2$ will be covered by the same capacity. The remaining $D_2$ can be met by non-shared renewable or fossil-based capacity. Shared fossil-based capacity is not considered, as $D_1$ is always met by cheaper renewable options in the first period. Thus, in the second period, the used options can be shared renewable, non-shared fossil, and non-shared renewable capacity. Note that this analysis assesses the affordability of different generation options but does not specify the prioritization or activation of specific options. The optimal choice depends on factors beyond cost, and the model automatically identifies the options used among the specified affordable options.

Figure \ref{fig:cluster 1} highlights the instance groups within cluster 1 based on the demand level in the second period ($D_2$) and the generation options that can be used in that period. The related instance groups are defined as follows:

\begin{figure}[htbp]
\centering
\begin{tikzpicture}
\begin{axis}[
  xlabel={\footnotesize $D_2$},
  xlabel style={at={(axis description cs:0.5,-0.07)}, anchor=north}, 
  ylabel={\footnotesize Average generation cost per unit},
  axis lines=left,
  xmin=0, xmax=14850,
  ymin=0, ymax=150,
  xtick={0,2000,4700,6000,10000,14000},
  ytick={0,31,61,102},
  xticklabels={{},$D_1$,$D_1+M_r$,$2M_r$,$2M_r+M_f$,$2M_r+2M_f$},
  yticklabels={{}, $\frac{CI_r}{2}+CP_r$,$CI_r+CP_r$, $CI_f+CP_f$},
  xticklabel style={font=\tiny},
  yticklabel style={font=\tiny},
  scaled x ticks=false,
  grid style=dashed,
  width=12.2cm,
  height=6cm,
  extra x ticks={3000},
  extra x tick labels={$M_r$}, 
  extra x tick style={grid style={draw=none}}
] 

\addplot[thick, fill=blue, fill opacity=0.2, draw=none] coordinates {(0,31) (2000,31)} \closedcycle;
\addplot[thick, fill=green, fill opacity=0.2,draw=none] coordinates {(2000,31) (2000,61) (4700,61)} \closedcycle;
\addplot[thick, fill=orange, fill opacity=0.2,draw=none] coordinates {(4700,61) (6000,61)} \closedcycle;
\addplot[thick, fill=yellow, fill opacity=0.2, draw=none] coordinates {(6000,101) (10000,101)} \closedcycle;
\addplot[thick, fill=red, fill opacity=0.2, draw=none] coordinates {(10000,101) (14000,101)} \closedcycle;
\addplot[thick] coordinates {(14000,101) (14800,101)};

\addplot[thick] coordinates {(0,31) (2000,31)};
\addplot[thick] coordinates {(2000,31) (2000,61)};
\addplot[thick] coordinates {(2000,61) (6000,61)};
\addplot[thick] coordinates {(6000,61) (6000,101)};
\addplot[thick] coordinates {(6000,101) (14000,101)};
\addplot[thick] coordinates {(14000,101) (14800,101)};

\addplot[red, dashed, thick] coordinates {(6000,0) (6000,120)}; 
\addplot[red, dashed, thick] coordinates {(14000,0) (14000,120)}; 

\node[font=\tiny] at (1000,15) {SR}; 
\node[font=\tiny] at (3350,30) {$R_2$};
\node[font=\tiny] at (5350,30) {$R_1$};
\node[font=\tiny] at (8000,50) {$F_2$};
\node[font=\tiny] at (12000,50) {$F_1$};

\node[font=\tiny, rotate=90] at (1000,52) {Ins. gr. 2};
\node[font=\tiny, rotate=90] at (3350,82) {Ins. gr. 3};
\node[font=\tiny, rotate=90] at (5350,82) {Ins. gr. 4};
\node[font=\tiny, rotate=45] at (6300,135) {Ins. gr. 5};
\node[font=\tiny, rotate=00] at (8000,108) {Ins. gr. 6};
\node[font=\tiny, rotate=00] at (12000,108) {Ins. gr. 7};
\node[font=\tiny, rotate=45] at (14300,135) {Ins. gr. 8};

\end{axis}
\end{tikzpicture}

\vspace{0em} 

\begin{tikzpicture}[every node/.style={rectangle, draw, thick, inner sep=2}]
\node[anchor=north west, font=\tiny, text width=11.8cm, outer sep=1pt, inner sep=2pt, draw, thick] at (6,0) {
\vspace{-7pt} 
\begin{tabbing}
\hspace{0.5cm}\= \hspace{5.5cm} \= \hspace{0.5cm} \= \kill
R1 \> Non-shared renewable capacity of period 1 \> R2 \> Non-shared renewable capacity of period 2\\
F1 \> Non-shared fossil-based capacity of period 1 \> F2 \> Non-shared fossil-based capacity of period 2\\
SR \> Shared renewable \\
\end{tabbing}
};
\end{tikzpicture}
\caption{Categorizing instance groups in cluster 1 based on generation options used for meeting second period demand}
\label{fig:cluster 1}
\end{figure}
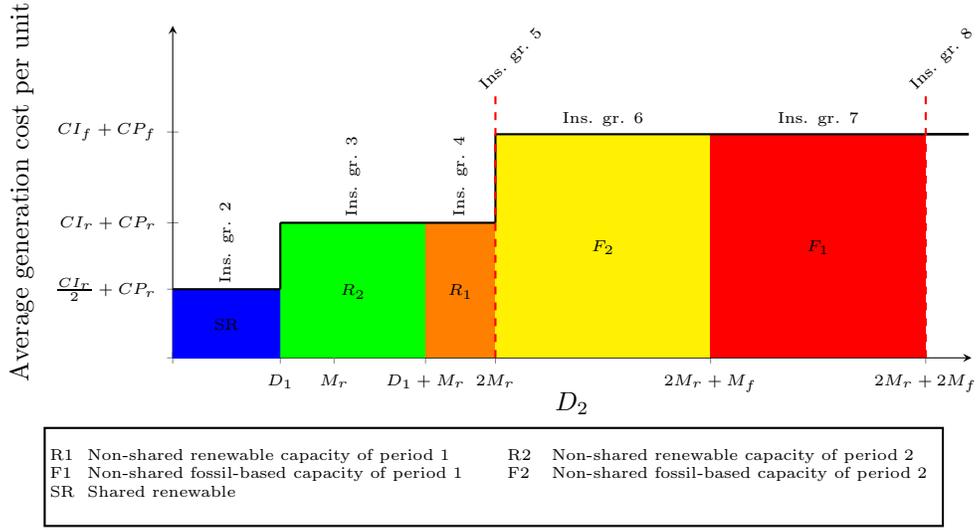

\begin{description}
    \item Instance group 1: $CL<\frac{CI_r}{2}+CP_r$, so no generation option is economically feasible. This leads to load shedding in both periods.
    
    \item Instance group 2: $\frac{CI_r}{2}+CP_r<CL$, so the only economically viable generation option is the shared renewable capacity. As a result, this capacity is used to meet $D_1$ units of the demand in the second period ($D_1$ units). The remaining demand experiences load shedding ($D_2-D_1$ units).

    \item Instance group 3: $CI_r+CP_r<CL$, so the economically viable options are the shared and non-shared capacities of renewables. Besides, $D_1<D_2<D_1+M_r$. Therefore, the non-shared renewable capacity emerges as the second cheapest option, capable of meeting up to $M_r$ units of the remaining demand in the second period following the use of the shared renewable capacity. 
    
    \item Instance group 4: Since $CI_r+CP_r<CL$, the shared and non-shared renewable capacities are economically viable options. Additionally, $D_1+M_r<D_2<2M_r$. Therefore, there is still unmet demand beyond the use of shared renewable capacity and non-shared renewable capacity of the second period. Consequently, the rest of $D_2$ is met by using the non-shared renewable capacity from the first period (up to $M_r-D_1$ units).
    
    \item Instance group 5: $CI_r+CP_r<CL$, so the shared and non-shared renewable capacities are affordable options; however, $D_2>2M_r$. Given that the renewable capacity is fully used over both periods and the use of fossil-based capacity is not cost-effective, any demand exceeding the capacity of renewable generators results in load shedding. ($D_2-2M_r$ units). 
    
    \item Instance group 6: Given that $CI_f+CP_f<CL$, the affordable options are shared and non-shared renewable capacities, as well as non-shared fossil-based capacity. However, since $2M_r<D_2<2M_r+M_f$, When $D_2$ exceeds the total renewable capacity across both periods, the non-shared fossil-based capacity of the second period is employed to meet the rest of $D_2$ (up to $M_f$ units).
    
    \item Instance group 7: $CI_f+CP_f<CL$, so the economically viable options are the shared and non-shared capacity of renewable and the non-shared capacity of fossil-based. Besides, $2M_r+M_f<D_2<2M_r+2M_f$. So, further investments in fossil-based generation in the first period are made just to meet the remaining demand in the second period (up to $M_f$ units).
    
    \item Instance group 8: Since $CI_f+CP_f<CL$, the affordable options are shared and non-shared renewable capacity and non-shared fossil-based capacity while $D_2>2M_r+2M_f$. In this case, renewable and fossil-based capacities during both periods reach their maximum investable capacity. However, there is unmet demand in the second period, resulting in load shedding.
\end{description}

Accordingly, 8 instance groups have been identified within Cluster 1. For each of these instance groups, there are different optimal decisions determined by the model. Additionally, after writing the complementary slackness conditions for the instance groups, specific values for LRMC for periods 1 and 2 in each instance group are obtained. The optimal decisions and LRMC values are summarized in Table \ref{tab:The optimal decisions and LRMCs for instance groups within cluster 1}. This table is related to Cluster 1 only. However, there are 6 clusters and 41 instance groups in total. If we write the complementary slackness conditions for each of the instance groups and find the related LRMC values, we can detect that only 7 distinct combinations of LRMC exist. These unique combinations of LRMCs are referred to as 'LRMC profiles.' All the instance groups, the obtained LRMC combinations, and their characteristics are displayed in \hyperref[sec:supplementaryTables]{Appendix A}, Table \ref{tab:instance_groups}. How the LRMC profiles are formed is studied in detail in the next subsection.  

\begin{table}[htbp]
\caption{The optimal decisions and LRMCs for instance groups within cluster 1}
\label{tab:The optimal decisions and LRMCs for instance groups within cluster 1}
\centering
\fontsize{7.5}{10}\selectfont
\scalebox{0.9}{
\begin{tabular}{p{0.3cm} >{\columncolor{gray!7}} p{1.7cm} p{1.3cm} >{\columncolor{gray!7}}p{0.8cm} p{1cm} >{\columncolor{gray!7}} p{1.1cm} p{0.8cm} >{\columncolor{gray!7}} p{0.8cm} p{0.4cm} >{\columncolor{gray!7}} p{0.8cm} p{1.45cm} >{\columncolor{gray!7}} p{1.45cm}}
\toprule
Ins. gr. & Demand                   & $CL$                     & Gen. option & $I_{r1}$  & $I_{r2}$  & $I_{f1}$       & $I_{f2}$   & $L_1$           & $L_2$           & $\lambda_1$     & $\lambda_2$ \\
\toprule
1          & $D_1<D_2$                & $CL<\frac{CI_r}{2}+CP_r$ & -           & $0$       & $0$       & $0$            & $0$        & $D_1$           & $D_2$           & $CL$            &$CL$         \\
\hline
2          & $D_1<D_2$                & $CL<CI_r+CP_r$           & SR          & $D_1$     & $0$       & $0$            & $0$        & $0$             & $D_2-D_1$       & $CI_r+2CP_r-CL$ & $CL$        \\
\hline
3          & $D_1<D_2<D_1+M_r$        & $CL<CI_r+CP_r$           & SR, R       & $D_1$     & $D_2-D_1$ & $0$            & $0$        & $0$             & $0$             & $CP_r$          & $CI_r+CP_r$ \\
\hline
4          & $D_1+M_r<D_2<2M_r$       & $CL<CI_r+CP_r$           & SR, R       & $D_2-M_r$ & $M_r$     & $0$            & $0$        & $0$             & $0$             & $CP_r$          & $CI_r+CP_r$ \\
\hline
5          & $D_2>2M_r$               & $CL<CI_f+CP_f$           & SR, R       & $M_r$     & $M_r$     & $0$            & $0$        & $0$             & $D_2-2M_r$      & $CP_r$          & $CL$        \\
\hline
6          & $2M_r<D_2<2M_r+M_f$      & $CL<CI_f+CP_f$           & SR, R, F    & $M_r$     & $M_r$     & $0$            & $D_2-2M_r$ & $0$             & $0$             & $CP_r$          & $CI_f+CP_f$ \\
\hline
7          & $2M_r+M_f<D_2<2M_r+2M_f$ & $CL<CI_f+CP_f$           & SR, R, F    & $M_r$     & $M_r$     & $D_2-2M_r-M_f$ & $M_f$      & $0$             & $0$             & $CP_r$          & $CI_f+CP_f$ \\
\hline
8          & $D_2>2M_r+2M_f$          & $CL<CI_f+CP_f$           & SR, R, F    & $M_r$     & $M_r$     & $M_f$          & $M_f$      & $0$             & $D_2-2M_r-2M_f$ & $CP_r$          & $CL$         \\
\bottomrule
\end{tabular}}
\end{table}

\subsection{LRMC profiles} \label{sec:LRMC profiles}

In this subsection, 7 LRMC profiles related to the instance groups are presented. Furthermore, it illustrates the conditions under which these profiles occur and investigates the cost recovery problem under the LRMC pricing method. In our analysis, we employ two arrow signs to convey specific pricing possibilities. $A \leftrightarrow B$ indicates that both prices can occur,i.e., price $A$ in the first period and price $B$ in the second period, as well as price $B$ in the first period and price $A$ in the second period. On the other hand, $A \rightarrow B$ signifies that only price $A$ occurs in the first period, and price $B$ takes place solely in the second period. Moreover, we denote off-peak demand period as $o$ and peak demand period as $p$, with demands in the off-peak and peak periods represented as $D_o$ and $D_p$, respectively. We can refer to Table \ref{tab:instance_groups} in the \hyperref[sec:supplementaryTables]{Appendix A} to see which LRMC profiles are related to which instance groups.

\begin{itemize}
    \item Profile 1: $CL \leftrightarrow CL$ 
\end{itemize}

In certain instance groups, load shedding occurs in both periods for various reasons. One situation is economic-based load shedding, which happens when it is more cost-effective to shed load than to generate electricity in both periods. Another instance is capacity-constrained load shedding, where there is not enough investable capacity to generate electricity despite its economic viability in both periods. Sometimes, a combination of these two types of load shedding can occur. In such instance groups, from a mathematical standpoint, the LRMCs are determined based on the following equations of the complementary slackness conditions:

\begin{equation}
\begin{cases}
\begin{aligned}
& L_1 \cdot (\lambda_1 - CL) = 0, \quad L_1>0\\
& L_2 \cdot (\lambda_2 - CL) = 0, \quad L_2>0\\
\end{aligned}
\end{cases}
\Rightarrow
\begin{cases}
\begin{aligned}
&\lambda_1 = CL \\
&\lambda_2 = CL \\
\end{aligned}   
\end{cases}
\end{equation}

\begin{itemize}
    \item Profile 2: $CL \leftrightarrow CI_g + 2CP_g - CL$
\end{itemize}

In such instance groups, off-peak demand ($D_o$) is fully met while the marginal generation option is the shared capacity of generator $g$; In the peak demand period, $D_o$ units of $D_p$ are satisfied by the same generation options used in the off-peak period. However, since the non-shard capacity of generator $g$ is not an affordable option, the remaining demand in the peak period ($D_p-D_o$) faces loadshed. The marginal option in the peak period becomes loadshed, and consequently, the LRMC during the peak period is equal to the loadshed cost ($CL$). On the other hand, investing in one more unit of shared capacity in the off-peak period will save one unit of loadshed in the peak-demand period. Therefore, the LRMC in the off-peak period will be the total generation cost of shared capacity during both period excluding the saved loadshed cost ($CI_g+2 \cdot CP_g-CL$).

From the mathematical point of view, the LRMCs are obtained as follows:
\begin{equation}
\begin{cases}
\begin{aligned}
&\beta_{gp} + \beta_{go} + \gamma_{gi'} = CI_g\\
&\lambda_p - \beta_{gp} = CP_g\\
&\lambda_o - \beta_{go} = CP_g\\
&\lambda_p = CL\\
&\gamma_{gi'} = 0 \\
\end{aligned}
\end{cases}
\Rightarrow
\begin{cases}
\begin{aligned}
&\lambda_o = CI_g+2CP_g-CL\\
&\lambda_p = CL
\end{aligned}   
\end{cases}
\end{equation}

\noindent where $i'=o$ if $o$ is the first period; otherwise, $i'=p$. Besides, generator $g$ is not generating at its maximum investable capacity, and so, $\gamma_{gi'}=0$. 



The cost recovery mechanism is examined through a comparative analysis of revenue and total generation costs over two periods. The objective is to demonstrate whether the revenue generated from supplying electricity meets or exceeds the total cost of the system, encompassing operational and investment expenses. In essence, this analysis seeks to establish that the financial profit of the power system is greater than or equal to zero. 

It is important to note that if the revenue is sufficient to cover the cost of the marginal generator, it will inherently cover the costs of less expensive generators, as their associated costs are lower. 
To further illustrate this point, consider the revenue derived from the marginal generator as follows:
\begin{align}
    &\text{Revenue} = D_o \cdot \lambda_o+D_p \cdot \lambda_p = D_o \cdot (CI_g+2CP_g-CL) + D_p \cdot CL \\
    & = D_o \cdot (CI_g+2CP_g) - D_o \cdot CL + D_p \cdot CL \notag \\
    & = D_o \cdot (CI_g+2CP_g)+(D_p-D_o) \cdot CL \notag
\end{align}
The total cost is:
\begin{align}
    & \text{Total cost} = D_o \cdot (CI_g+2CP_g)+(D_p-D_o) \cdot CL  
\end{align}
Here, $D_o$ units in both periods are satisfied by the electricity generated by shared capacity, and $D_o \cdot (CI_g + 2CP_g)$ represents the generation cost for the shared capacity used in both periods. In the peak period, $D_p - D_o$ units face load shedding, and $(D_p - D_o) \cdot CL$ represents the cost of load shedding. 

Now, we calculate the profit as follows:
\begin{align}
    & \text{Profit = Revenue - Total cost} \\
  =& D_o \cdot (CI_g+2CP_g)+(D_p-D_o) \cdot CL \notag\\
  -& (D_o \cdot (CI_g+2CP_g)+(D_p-D_o) \cdot CL) = 0 \notag
\end{align}
The analysis concludes that the revenue from electricity generation is sufficient to cover the total costs.
\begin{itemize}
    \item Profile 3: $CP_g \leftrightarrow CI_g + CP_g$
\end{itemize}

The LRMC profile indicates that during the off-peak period, the shared capacity of generator $g$ is the marginal option, while in the peak period, the non-shared capacity of generator $g$ is the marginal option. In the peak period, the LRMC is $CI_g + CP_g$ because meeting additional demand requires investing in non-shared capacity, which is only used during this period. Conversely, in the off-peak period, the LRMC is $CP_g$ because an increase in demand is met by shared capacity, which incurs an investment cost of $CI_g + 2CP_g$ during both periods. However, this is offset by a reduction in the need for non-shared capacity during the peak period, effectively making the LRMC for the off-peak period $CI_g+2CP_g - (CI_g+CP_g)=CP_g$.

From a mathematical perspective, the LRMCs are derived using the following approach: 

\begin{equation}
\begin{cases}
\begin{aligned}
&\beta_{gp} + \beta_{go} + \gamma_{gi'} = CI_g\\
&\lambda_p - \beta_{gp} = CP_g\\
&\lambda_o - \beta_{go} = CP_g\\
&\gamma_{gi'} = 0 \\
\end{aligned}
\end{cases}
\Rightarrow
\begin{cases}
\begin{aligned}
&\lambda_{gp} + \lambda_{go} = CI_g + 2CP_g\\
&\lambda_p - \beta_{gp} = CP_g\\
&\lambda_o - \beta_{go} = CP_g\\
\end{aligned}   
\end{cases}
\end{equation}
where $i'=o$ if $o$ is the first period; otherwise, $i'=p$. Moreover, generator $g$ is not generating at its maximum investable capacity, and so, $\gamma_{gi'}=0$. According to $\beta$, two situations can occur: when the peak period is the first period ($p=1, o=2$) and when the peak period is the second period ($o=1, p=2$). When the peak period is the first period, the invested capacity in the first period is sufficient to meet all the demand in the second period. Consequently, all the invested capacities during the first and second periods are higher than the generated electricity in the second period, i.e., $P_{go}<I_{gp}+I_{go}$; therefore, $\beta_{go}=0$, and this leads to:
\begin{equation}
\begin{cases}
\begin{aligned}
&\lambda_p= CI_g+CP_g\\
&\lambda_o=CP_g\\
\end{aligned}
\end{cases}
\end{equation}
On the other hand, if the peak period is in the second period ($o=1, p=2$), the last unit generated in the second period using the non-shared capacity becomes the decisive factor in determining the cost distribution. In this case, $I_{gp}>0$, leading to $\beta_{gp}+\gamma_{gp}=CI_g$. Since $\gamma_{gp}=0$, we will have $\beta_{gp}=CI_g$. Finally, LRMCs will be as follows:

\begin{equation}
\begin{cases}
\begin{aligned}
&\lambda_o=CP_g\\
&\lambda_p= CI_g+CP_g\\
\end{aligned}
\end{cases}
\end{equation}

For evaluating the cost recovery, consider the following:
\begin{align}
     \text{Revenue} =&  D_o \cdot CP_g + D_p \cdot (CI_g+CP_g) \\
                     =& D_o \cdot CP_g + D_o \cdot (CI_g+CP_g) + (D_p-D_o) \cdot (CI_g+CP_g) \notag \\
                     =& D_o \cdot (CI_g+2CP_g)+(D_p-D_o) \cdot (CI_g+CP_g) \notag  
\end{align}
\begin{align}
    & \text{Total cost} = D_o \cdot (CI_g+2CP_g)+(D_p-D_o) \cdot (CI_g+CP_g) \label{eq:Generation cost P3}
\end{align}
In Equation \ref{eq:Generation cost P3} , the $D_o$ units in both periods are supplied by the shared capacity, with an associated generation cost of $D_o \cdot (CI_g + 2CP_g)$. The remaining $D_p - D_o$ units in the peak period are supplied by non-shared capacity, with a generation cost of $(D_p - D_o) \cdot (CI_g + CP_g)$. Finally, the profit is calculated as follows:
\begin{align}
     \text{Profit} =& D_o \cdot (CI_g+2CP_g)+(D_p-D_o) \cdot (CI_g+CP_g) \\
  - & (D_o \cdot (CI_g+2CP_g)+(D_p-D_o) \cdot (CI_g+CP_g)) = 0 \notag
\end{align}
Therefore, the revenue and total cost are equal, indicating that the total cost is recovered during both periods. 


\begin{itemize}
    \item Profile 4: $CL \leftrightarrow CP_g$
\end{itemize}

This LRMC profile occurs when the shared capacity of generator type $g$ meets all demand during the off-peak period. However, in the peak period, the shared capacity is insufficient. The non-shared capacity of both peak and off-peak periods is then used to meet the additional peak demand, reaching its maximum investable capacity. Despite these efforts, some peak demand remains unmet and faces load shedding. Therefore, the LRMC during the peak period is $CL$. On the other hand, the LRMC in the off-peak period is $CP$ due to the existence of additional invested capacity in this period. This idle investment capacity remains unused during the off-peak period, and an increase in demand by a unit can be satisfied without further investment, entailing only operational cost.

Mathematically speaking, the LRMCs are calculated as follows:

\begin{equation}
\begin{cases}
\begin{aligned}
&\beta_{go} = 0\\
&\lambda_o - \beta_{go} = CP_g\\
&\lambda_p = CL\\
\end{aligned}
\end{cases}
\Rightarrow
\begin{cases}
\begin{aligned}
&\lambda_o = CP_g\\
&\lambda_p = CL
\end{aligned}   
\end{cases}
\end{equation}
As a result of the additional investment during the off-peak period, we have $P_{go}<I_{go}+I_{gp}$. This leads to the deduction that the associated dual variable becomes zero ($\beta_{go} = 0$), and subsequently $\lambda_o=CP_g$. Furthermore, with load shedding in the peak period, we have $\lambda_p = CL$.

Now, our goal is to investigate the cost recovery for this case. Consider the total cost as follows:
\begin{align}
    & \text{Total cost} \\
    =& D_o \cdot (CI_g+2CP_g) + N \cdot (CI_g+CP_g) + (D_p-D_o-N) \cdot  CL \notag
\end{align}
The shared capacity of generator $g$ is used during both peak and off-peak periods, with a generation cost of $D_o \cdot (CI_g + 2CP_g)$. In the peak period, the remaining demand $N$ is met by non-shared capacity at a cost of $N \cdot (CI_g + CP_g)$, while any additional demand faces load shedding.

The revenue is given by: 
\begin{align}
     \text{Revenue}=& D_o \cdot CP_g + D_p \cdot CL = D_o \cdot CP_g+ (D_o+N+(D_p-D_o-N)) \cdot CL \notag \\
    =& D_o \cdot (CP_g+CL) + N \cdot CL + (D_p-D_o-N) \cdot CL
\end{align}
Given that $CL > CI_g + CP_g$, the profit can be calculated as follows:
\begin{align}
    & \text{Profit}= \\
    & D_o \cdot (CP_g+CL) + N \cdot CL + (D_p-D_o-N) \cdot CL  \notag\\
    -& (D_o \cdot (CI_g+2CP_g) + N \cdot (CI_g+CP_g) + (D_p-D_o-N) \cdot  CL)= \notag\\
    & D_o \cdot (CL-(CI_g+CP_g)) + N \cdot (CL-(CI_g+CP_g))>0 \notag
\end{align}
This calculation demonstrates that the total cost is lower than the revenue, showing that the cost can be recovered by the revenue.
\begin{itemize}
    \item Profile 5: $CL \rightarrow CI_g + CP_g$
\end{itemize}

In this LRMC profile, demand levels during both periods exceed $M_r+M_f$. Consequently, the generation from the shared capacity of the renewable and fossil-based generator reaches its maximum investable capacity in the first period. However, due to insufficient capacity to meet the entire demand in the first period, load shedding occurs, leading to an LRMC value of $CL$. In the second period, the remaining demand is satisfied using the non-shared capacity of generator $g$, and it operates below its maximum investable capacity. Therefore, if there is extra demand during this period, it should be met by investing in new capacity for generator $g$. This yields an LRMC of $CI_g+CP_g$. From a mathematical standpoint, the LRMC profile is derived as follows:

\begin{equation}
\begin{cases}
\begin{aligned}
&\lambda_1 = CL\\
&\lambda_2 - \beta_{g2} = CP_g\\
&\beta_{g2} - \gamma_{g2} = CI_g\\
&\gamma_{g2} = 0 \\
\end{aligned}
\end{cases}
\Rightarrow
\begin{cases}
\begin{aligned}
&\lambda_1 = CL\\
&\lambda_2 = CI_g+CP_g
\end{aligned}   
\end{cases}
\end{equation}

Concerning cost recovery, the total cost is as:
\begin{align}
     \text{Total cost} & =  (M_r+M_f) \cdot (CI_g+2CP_g) + (D_1-(M_r+M_f)) \cdot CL \notag \\
    & + N \cdot (CI_g+CP_g)
\end{align}
Here, $N=D_2-(M_r+M_f)$ represents the units met by non-shared capacity in the second period. $D_1$ units of demand in each of the two periods are satisfied by shared capacity, and the cost of generating these units is $(M_r + M_f) \cdot (CI_g + 2CP_g)$. The generation cost of the remaining units met in the second period is $N \cdot (CI_g + CP_g)$. The loadshed cost in the first period is addressed by $(D_1 - (M_r + M_f)) \cdot CL$.

Revenue can be written as:
\begin{align}
    &\text{Revenue} = D_1 \cdot CL+D_2 \cdot (CI_g+CP_g) \\
    =&(M_r+M_f+(D_1-(M_r+M_f))) \cdot CL + (M_r+M_f+N) \cdot (CI_g+CP_g) \notag \\
    =&(M_r+M_f) \cdot CL + (D_1-(M_r+M_f)) \cdot CL \notag \\ 
    +&(M_r+M_f) \cdot (CI_g+CP_g) + N \cdot (CI_g+CP_g) \notag \\
    =&(M_r+M_f) \cdot (CL+CI_g+CP_g) + (D_1-(M_r+M_f)) \cdot CL + N \cdot (CI_g+CP_g) \notag
\end{align}
Given that $CL > CI_g + CP_g$, the profit will be as: 
\begin{align}
    & \text{Profit} \\
    =&(M_r+M_f) \cdot (CL+CI_g+CP_g) + (D_1-(M_r+M_f)) \cdot CL + N \cdot (CI_g+CP_g) \notag \\
    -&((M_r+M_f) \cdot (CI_g+2CP_g) + (D_1-(M_r+M_f)) \cdot CL + N \cdot (CI_g+CP_g)) \notag \\
    =& (M_r+M_f) \cdot (CL-CP_g)>0 \notag
\end{align}
In essence, the revenue is high enough to ensure that all associated costs, including those for generation and load shedding, are recovered.

\begin{itemize}
    \item Profile 6: $CP_r \leftrightarrow CI_f + CP_f$
\end{itemize}

This LRMC occurs in two situations. First, when $D_1 > D_2$, such that $D_2 < M_r$ and $M_r < D_1 < M_r + M_f$. In this case, the marginal option in the second period is the shared capacity of renewables. Additionally, non-shared renewable and fossil-based capacities are used in the first period, making the LRMC $CI_f + CP_f$, as an increase in peak demand requires more fossil-based investment. In the second period, if demand increases by one unit, the cost of meeting this demand is $CP_r$ because of the higher investment in the first period, resulting in idle capacity in the second period. This idle capacity is used to meet increased demand, with the only cost being $CP_r$. Mathematically, the subsequent computations arise:
\begin{equation}
\begin{cases}
\begin{aligned}
&\gamma_{f1} = 0\\
&\beta_{f2} = 0\\
&\lambda_1 - \beta_{f1} = CP_f\\
&\beta_{f1} + \beta_{f2} - \gamma_{f1} = CI_f\\
&\beta_{r2}=0\\
&\lambda_2 - \beta_{r2} = CP_r\\
\end{aligned}
\end{cases}
\Rightarrow
\begin{cases}
\begin{aligned}
&\lambda_1 = CI_f+CP_f\\
&\lambda_2 = CP_r
\end{aligned}   
\end{cases}
\end{equation}
In this situation, because of idle capacity in the second period, we will have $\beta_{f2}=0$ and $\beta_{r2}=0$ (i.e., $P_{r2}<I_{r1}+I_{r2} \rightarrow \beta{r2}=0 $ and $P{f2}<I_{f1}+I_{f2} \rightarrow \beta{f2}=0$). Furthermore, considering that the fossil-based generator acts as the marginal generator in the first period, although it is not operating at its maximum investable capacity, it follows that $\gamma_{f1}=0$. Finally, according to the calculations, $\lambda_1=CI_f+CP_f$ and $\lambda_2=CP_r$. 

In the second situation, $D_1 < D_2$, with $D_1 < M_r$ and $2M_r < D_2 < 2M_r + 2M_f$. The marginal option in the first period is shared renewable capacity. The second period uses shared renewable capacity, non-shared renewable capacity from both periods, and non-shared fossil capacity from the second period, with the latter also used in the first period if needed. Consequently, the LRMC in the second period is $CI_f + CP_f$, while the LRMC in the first period is $CP_r$ due to the extra renewable investment. Mathematically speaking, the subsequent calculations come into play:

\begin{equation}
\begin{cases}
\begin{aligned}
&\beta_{r1} = 0\\
&\lambda_1 - \beta_{r1} = CP_r\\
&\gamma_{f2} = 0\\
&\beta_{f2} - \gamma_{f2} = CI_f\\
&\lambda_2 - \beta_{f2} = CP_f\\
\end{aligned}
\end{cases}
\Rightarrow
\begin{cases}
\begin{aligned}
&\lambda_1 = CP_r\\
&\lambda_2 = CI_f+CP_f
\end{aligned}   
\end{cases}
\end{equation}
In this situation, $P_{r1} < I_{r1} = M_r \rightarrow \beta_{r1} = 0$. Meanwhile, in the subsequent period, the fossil-based generator is not operating at its maximum investable capacity, thus yielding $\gamma_{f2} = 0$. The calculations conclusively reveal that the LRMC for the first period is $CP_r$, while for the second period, it is $CI_f + CP_f$.

For examining cost recovery, consider the total cost as follows:
\begin{align}
    \text{Total cost}=& D_o \cdot (CI_r+2CP_r)+ N \cdot (CI_r+CP_r) \\
                          +& (D_p-D_o-N) \cdot (CI_f+CP_f) \notag 
\end{align}
Here, $N$ is the portion of $D_p$ met by non-shared renewable capacity. The total cost includes the costs of using shared renewable capacity in both periods, $D_o \cdot (CI_r + 2CP_r)$, non-shared renewable capacity during the peak period, $N \cdot (CI_r + CP_r)$, and non-shared fossil capacity during the peak period, $(D_p - D_o - N) \cdot (CI_f + CP_f)$.

The revenue can be expressed as:
\begin{align}
     & \text{Revenue}=D_o \cdot CP_r + D_p \cdot (CI_f+CP_f) \\
    =& D_o \cdot CP_r + (D_o + N + (D_p-D_o-N)) \cdot (CI_f+CP_f) \notag\\
    =& D_o \cdot (CP_r+CI_f+CP_f) + N \cdot (CI_f+CP_f) + (D_p-D_o-N) \cdot (CI_f+CP_f) \notag
\end{align}

Given that $CI_f + CP_f > CI_r + CP_r$, the profit is calculated as follows:
\begin{align}
    &\text{Profit}= \\
    & D_o \cdot (CP_r+CI_f+CP_f) + N \cdot (CI_f+CP_f) + (D_p-D_o-N) \cdot (CI_f+CP_f) \notag\\
   -& (D_o \cdot (CI_r+2CP_r)+ N \cdot (CI_r+CP_r) + (D_p-D_o-N) \cdot (CI_f+CP_f)) \notag\\
   =& D_o \cdot ((CI_f+CP_f)-(CI_r+CP_r)) + N \cdot ((CI_f+CP_f)-(CI_r+CP_r))>0 \notag
\end{align}
So, the total cost is recovered by the revenue.

\begin{itemize}
    \item Profile 7: $CI_f+2CP_f-(CI_r+CP_r) \rightarrow CI_r+CP_r$
\end{itemize}

This LRMC profile occurs when $D_1 < D_2$, such that $M_r < D_1 < M_r + M_f$ and $D_1 < D_2 < D_1 + M_r$. All renewable capacity of the first period is invested, reaching its maximum investable capacity, and the remaining $D_1 - M_r$ units are met by fossil-based capacity, not reaching its maximum. These capacities are utilized in the second period as shared capacity. The remaining $D_2 - D_1$ units in the second period are met by invested renewable capacity, used only in this period. Thus, the LRMC for the second period is $CI_r + CP_r$. If demand in the first period increases by one unit, new investment in fossil-based capacity is required, which will also meet one unit of demand in the second period, saving the need for one unit of non-shared renewable capacity. So, the generation cost of one unit by the newly invested capacity in both periods will be $CI_f + 2CP_f$, which saves a cost of $CI_r + CP_r$. Therefore, the LRMC of the first period is $CI_f + 2CP_f - (CI_r + CP_r)$. From a mathematical perspective, the LRMCs are computed as follows:
\begin{equation}
\begin{cases}
\begin{aligned}
&\lambda_2 - \beta_{r2} = CP_r \\
&\beta_{r2} - \gamma_{r2} = CI_r \\
&\gamma_{r2} = 0 \\
&\beta_{f1} + \beta_{f2} - \gamma_{f1} = CI_f\\
&\gamma_{f1} = 0\\
&\lambda_2 - \beta_{f2} = CP_f \\
\end{aligned}
\end{cases}
\Rightarrow
\begin{cases}
\begin{aligned}
&\lambda_1 = CI_f+2CP_f-(CI_r+CP_r)\\
&\lambda_2 = CI_r+CP_r
\end{aligned}   
\end{cases}
\end{equation}

Regarding cost recovery, consider the total cost as:
\begin{align}
    \text{Total cost}=& M_r \cdot (CI_r+2CP_r) + (D_1-M_r) \cdot (CI_f+2CP_f) \notag \\
    +& (D_2-D_1) \cdot (CI_r+CP_r) 
\end{align}
Here, $M_r \cdot (CI_r + 2CP_r)$ represents the generation cost from shared renewable capacity, $(D_1 - M_r) \cdot (CI_f + 2CP_f)$ represents the generation cost from shared fossil capacity, and $(D_2 - D_1) \cdot (CI_r + CP_r)$ is the generation cost from non-shared renewable capacity in the second period.

The revenue generated from the supplied electricity is as follows:
\begin{align}
    & \text{Revenue}= D_1 \cdot (CI_f+2CP_f-(CI_r+CP_r)) + D_2 \cdot (CI_r+CP_r) \\
    = & (M_r+(D_1-M_r)) \cdot (CI_f+2CP_f-(CI_r+CP_r)) \notag \\
    + & (M_r+(D_1-M_r)+(D_2-D_1)) \cdot (CI_r+CP_r) \notag \\
    = & M_r \cdot (CI_f+2CP_f) + (D_1-M_r) \cdot (CI_f+2CP_f) + (D_2-D_1) \cdot (CI_r+CP_r) \notag
\end{align}
Knowing that $CI_f + 2CP_f > CI_r + 2CP_r$, the profit is calculated as follows:
\begin{align}
    & \text{Profit}= \\
    & M_r \cdot (CI_f+2CP_f) + (D_1-M_r) \cdot (CI_f+2CP_f) + (D_2-D_1) \cdot (CI_r+CP_r) \notag \\
   -& (M_r \cdot (CI_r+2CP_r) + (D_1-M_r) \cdot (CI_f+2CP_f) + (D_2-D_1) \cdot (CI_r+CP_r)) \notag \\
   =& M_r \cdot ((CI_f+2CP_f)-(CI_r+2CP_r))>0 \notag
\end{align}
The revenue exceeds the total cost. Therefore, cost recovery is assured.

\subsection{Implications}

In conclusion, the aforementioned profiles offer insights into the allocation of LRMCs across distinct periods, each depending on various conditions. Notably, cost recovery remains assured for all LRMC profiles under the LRMC pricing method, despite the variations in cost allocation mechanisms. Observations show that if a generator is invested in a period and used solely in that period, the total investment cost of that generator is assigned to that period (known as non-shared capacity). However, if a generator is invested in one period and used in both peak and off-peak periods (referred to as shared capacity), the investment cost is distributed across both periods in different ways, depending on the LRMC profile. In total, the cost recovery occurs over both periods.  

One distribution pattern is to assign all investment costs of shared capacity exclusively to the peak period. Consider that the marginal generation option in the off-peak period is the shared capacity of generator $g$, and the marginal option in the peak period is the non-shared capacity of generator $g'$. The pattern occurs if the generator $g'$ is the same as or more expensive than generator $g$ in terms of investment cost ($CI_g \leq CI_{g'}$), the total investment cost of the shared capacity is allocated to the peak period. Otherwise ($CI_g > CI_{g'}$), the investment cost of $CI_g$ is distributed such that $CI_{g'}$ is allocated to the peak period, and the remaining $CI_g - CI_{g'}$ is allocated to the off-peak period.

The same rule applies when load shedding occurs in the peak period. If $CI_g \leq CL$, the total investment cost is allocated to the peak period. Otherwise ($CI_g > CL$), the investment cost of $CI_g$ is distributed such that from $CI_g$, $CL$ is allocated to the peak period, and the remaining $CI_g - CL$ is allocated to the off-peak period.

\section{SRMC estimation methodology} \label{sec:SRMC estimation methodology}

In this section, our objective focus is on the SRMC model. To formulate the SRMC model, we start by setting the investment capacity values to their optimal levels determined by the LRMC model. This procedure establishes the model referred to as the 'primal model for SRMC'. The next step explores the dual of the flow-balance constraints within this model to determine the SRMCs. In this context, we assume the optimal investment capacity values to be denoted as $I^*$. Consequently, the primal model for SRMC can be formulated as follows:
\begin{align}
    \min z &= CI_r \cdot I^*_{r1} + CI_r \cdot I^*_{r2} + CI_f \cdot I^*_{f1} + CI_f \cdot I^*_{f2} + CP_r \cdot P_{r1} + CP_r \cdot P_{r2} \notag \\ 
    &\quad +  CP_f \cdot P_{f1} + CP_f \cdot P_{f2} + CL \cdot L_1 + CL \cdot L_2 && 
\end{align}
\begin{alignat*}{2}
    s.t.  &\quad P_{r1} + P_{f1} + L_1 = D_1 && \quad (\dot{\lambda_1} \in (-\infty, +\infty)) \\
          &\quad P_{r2} + P_{f2} + L_2 = D_2 && \quad (\dot{\lambda_2} \in (-\infty, +\infty)) \\
          &\quad -P_{r1} \geq -I^*_{r1} && \quad (\dot{\beta_{r1}} \geq 0)\\
          &\quad -P_{r2} \geq -I^*_{r1} - I^*_{r2}  && \quad (\dot{\beta_{r2}} \geq 0) \\
          &\quad -P_{f1} \geq -I^*_{f1}  && \quad (\dot{\beta_{f1}} \geq 0)     \\  
          &\quad -P_{f2} \geq -I^*_{f1} - I^*_{f2}  && \quad (\dot{\beta_{f2}} \geq 0) \\
          &\quad P \geq 0; L \geq 0 
\end{alignat*}

In this model, the constraints pertaining to investment capacity limitations have been excluded since investment capacities are parameters. Consequently, the associated dual variables are omitted. Moreover, the dual variables of the constraints in the SRMC model are represented by dotted symbols. Now, the dual model for SRMC is obtained as follows:
\begin{align}
    \max y &= CI_r \cdot I^*_{r1} + CI_r \cdot I^*_{r2} + CI_f \cdot I^*_{f1} + CI_f \cdot I^*_{f2} + D_1 \cdot \dot{\lambda_{1}} + D_2 \cdot \dot{\lambda_{2}} \notag\\ 
    &\quad - I^*_{r1} \cdot \dot{\beta_{r1}} - (I^*_{r1} + I^*_{r2}) \cdot \dot{\beta_{r2}} - I^*_{f1} \cdot \dot{\beta_{f1}} - (I^*_{f1} + I^*_{f2}) \cdot \dot{\beta_{f2}} &&
\end{align}
\vspace{-2em}
\begin{alignat*}{2}
    s.t.  &\quad \dot{\lambda_{1}} - \dot{\beta_{r1}} \leq CP_r && \quad (P_{r1} \geq 0) \\
          &\quad \dot{\lambda_{2}} - \dot{\beta_{r2}} \leq CP_r && \quad (P_{r2} \geq 0) \\
          &\quad \dot{\lambda_{1}} - \dot{\beta_{f1}} \leq CP_f && \quad (P_{f1} \geq 0) \\
          &\quad \dot{\lambda_{2}} - \dot{\beta_{f2}} \leq CP_f && \quad (P_{f2} \geq 0) \\
          &\quad \dot{\lambda_1} \leq CL && \quad (L_1 \geq 0)\\
          &\quad \dot{\lambda_2} \leq CL && \quad (L_2 \geq 0)\\
          &\quad \dot{\beta} \geq 0; \dot{\lambda} \in (-\infty, +\infty)
\end{alignat*}

Now, our objective is to determine the value of the SRMCs. Similar to the approach employed in the computation of LRMCs, we rely on the principles of complementary slackness to achieve this goal. The instance groups discussed earlier for LRMCs also apply to SRMCs. By deriving the complementary slackness conditions for all instance groups, we found that the relationship between LRMC and SRMC varies under specific conditions: when LRMC in a period equals CL, when it equals CP, and when it lies between CP and CL.

\begin{itemize}
    \item $\lambda=CL \Rightarrow \dot{\lambda}=CL$
\end{itemize}

This is a situation where loadshed occurs in a period, and therefore, the LRMC for this period becomes equivalent to the loadshed cost; i.e., $\lambda=CL$. Since the optimal solution is the same for both LRMC and SRMC models, if load shedding occurs in a period in the LRMC model, it will also occur in the SRMC model. Consequently, in the primal model for SRMC, we have: 
\begin{alignat}{3}
&L \cdot (\dot{\lambda} - CL) = 0 : &&\quad L>0  &&\quad \Rightarrow \dot{\lambda} = CL
\end{alignat}
According to this condition, the SRMC for this period equals loadshed cost. To conclude, when load shedding occurs during a period, both LRMC and SRMC will be equivalent, both amounting to the loadshed cost ($\lambda=\dot{\lambda}=CL$). 

\begin{itemize}
    \item $\lambda=CP \Rightarrow \dot{\lambda}=CP$
\end{itemize}

In this case, the LRMC in a period equals the operational cost. This occurs under two conditions. The first condition is when there is additional investment in the off-peak period because it is necessary to meet the demand in the peak period. Consequently, the generation in the off-peak period will be less than the invested capacity. Given that the optimal solution for both the primal models of LRMC and SRMC is equivalent, the additional generation observed within the long-run context also translates to the SRMC model. Therefore, in the SRMC model, we have $P_{go} \leq I^*_{go} + I^*_{gp}$. As a result, the associated dual variable becomes zero ($\dot{\beta_{go}} = 0$). With the presence of generation in this period ($P_{go} \geq 0$), and according to the complementary slackness conditions, this relationship holds: $\dot{\lambda_{o}} - \dot{\beta_{go}} = CP_g$, leading to $\dot{\lambda_o} = CP_g$. This implies that when $\lambda_o = CP_g$, we have $\dot{\lambda_o} = CP_g$.

The second condition occurs when the marginal option is the shared capacity of generator $g$ in the off-peak period and the non-shared capacity of that generator in the peak period. In this case, there is investment and generation in both periods, while no load shedding occurs. So, the complementary slackness conditions of the SRMC can be formed as follows:
\begin{align}
&P_{go} \cdot (\dot{\lambda_{o}} - \dot{\beta_{go}} - CP_g) = 0 : &&\quad P_{go}>0 &&\quad \Rightarrow \dot{\lambda_o} - \dot{\beta_{go}} = CP_g \notag\\
&P_{gp} \cdot (\dot{\lambda_{p}} - \dot{\beta_{gp}} - CP_g) = 0 : &&\quad P_{gp}>0 &&\quad \Rightarrow \dot{\lambda_p} - \dot{\beta_{gp}} = CP_g\notag\\
&L_o \cdot (\dot{\lambda_o} - CL) = 0 : &&\quad L_o=0  &&\quad \Rightarrow \dot{\lambda_o} \leq CL\notag\\
&L_p \cdot (\dot{\lambda_p} - CL) = 0 : &&\quad L_p=0  &&\quad \Rightarrow \dot{\lambda_p} \leq CL\\
&\dot{\beta_{go}} \cdot (-P_{go} + I^*_{go}) = 0 : &&\quad (P_{go})=(I^*_{go}) &&\quad \Rightarrow \dot{\beta_{go}} \geq 0 \notag\\
&\dot{\beta_{gp}} \cdot (-P_{gp} + I^*_{go} + I^*_{gp}) = 0 : &&\quad(P_{gp})=(I^*_{go}+I^*_{gp}) &&\quad \Rightarrow \dot{\beta_{gp}} \geq 0 \notag
\end{align}
Ultimately, based on these conditions, we conclude that $CP_g \leq \dot{\lambda_o} \leq CL$.

In this particular case, when the LRMC equals the operational cost, the SRMC can range between the operational cost and the loadshed cost. This occurs because the SRMC's coefficient in the objective function of the dual model is zero, allowing SRMC to assume multiple values. This phenomenon is a result of degeneracy in the primal model, where redundant constraints or the problem's structure lead to multiple equivalent optimal solutions in the dual model. As a result, the value of the dual variable fails to provide reliable interpretations for SRMCs. To resolve this degeneracy, we add a small value to the invested capacity values ($I^*$), which prevents a zero coefficient for SRMC in the dual objective function. For further details on the causes of degeneracy and the resolution method, please refer to \hyperref[sec:degeneracy]{Appendix B}. After resolving the degeneracy, the SRMC takes the lowest value within the interval $[CP_g, CL]$, which is $CP_g$. In summary, it is demonstrated that when $\lambda_{o} = CP_g$, $\dot{\lambda_{o}}$ also equals $CP_g$.

\begin{itemize}
    \item $CP<\lambda<CL \Rightarrow \dot{\lambda}=CP$
\end{itemize}

In the primal model for LRMC, such cases happen when there is no additional investment in period $t$, and neither is there load shedding during that period.  If we suppose that generator $g$ is the marginal generator in period $t$; so, $\lambda_t \neq CP_g,  \lambda_t \neq CL $. Therefore, the LRMC in period $t$ will take a value between $CP_g$ and $CL$ ($CP_g<\lambda_t<CL$). 

Acknowledging the same optimal solution for the LRMC and SRMC models, it follows that cases, when no additional investment or load shedding occurs in the LRMC model, are reflected similarly in the SRMC model. Consequently, when additional generation is absent in the SRMC model for period $t$, we have:
\begin{alignat}{3}
&\dot{\beta_{gt}} \cdot (-P_{gt} + I^*_{gt}) = 0 : &&\quad (P_{gt})=(I^*_{gt}) &&\quad \Rightarrow \dot{\beta_{gt}} \geq 0 
\end{alignat}
On the other hand, there is no loadshed here; therefore:
\begin{alignat}{3}
&L_t \cdot (\dot{\lambda_t} - CL) = 0 : &&\quad L_t=0  &&\quad \Rightarrow \dot{\lambda_t} \leq CL
\end{alignat}

The demand in a given period can be satisfied either through generated electricity or, in its absence, by facing load shedding. Note that in the present context, load shedding is not a factor. Thus, the demand must be met by electricity generation. Therefore, we will have $P_{gt} > 0$, assuming that $g$ is the marginal generator type in this period. This observation leads to the following complementary slackness condition:
\begin{alignat}{3}
&P_{gt} \cdot (\dot{\lambda_t} - \dot{\beta_{gt}} - CP_g) = 0 : &&\quad P_{gt}>0 &&\quad \Rightarrow \dot{\lambda_t} - \dot{\beta_{gt}} = CP_g
\end{alignat}
Based on these three relationships, we can derive the following:
\begin{equation}
\begin{cases}
\begin{aligned}
&\dot{\beta_{gt}} \geq 0\\
&\dot{\lambda_t} \leq CL\\
&\dot{\lambda_t} - \dot{\beta_{gt}} = CP_g\\
\end{aligned}
\end{cases}
\Rightarrow
\begin{cases}
\begin{aligned}
& CP_g \leq \dot{\lambda_t} \leq CL
\end{aligned}   
\end{cases}
\end{equation}
Hence, when the LRMC of a period falls between the operational cost and the loadshed cost, the corresponding SRMC can be any value within the $[CP_g, CL]$ interval. As discussed earlier, this multiplicity of optimal solutions in the dual model for SRMC occurs when the primal SRMC encounters degeneracy. To address this, resolving the degeneracy becomes imperative. After resolution, we will have $\dot{\lambda} = CP$. In summary, it can be concluded that when $CP < \lambda < CL$ in a period, the SRMC in that period is equal to $CP$.

\subsection{Implications}

The culmination and findings of these three subsections are summarized as follows:
\begin{enumerate}
    \item $ \lambda_t = CP_{g} \rightarrow \dot{\lambda_t} = CP_{g} $ 
    \item $ \lambda_t = CL \rightarrow \dot{\lambda_t} = CL $
    \item $ CP_{g} < \lambda_t < CL \rightarrow \dot{\lambda_t} = CP_{g} $
\end{enumerate}
The results underscore the significance of addressing degeneracy in the primal model of SRMC, as its consideration is pivotal for ensuring the determination of a reliable SRMC. Based on the results, the SRMC includes only operational cost ($CP$) or load shedding cost ($CL$), which means that the calculated cost for generating an additional unit of electricity considers only the immediate costs associated with operating existing facilities or, in the case of load shedding, the cost of not supplying electricity to meet demand. This allows for a more dynamic and flexible assessment of costs in response to short-term changes in demand or operational conditions. 

Moreover, based on our findings, we observe a consistent trend: SRMC values consistently remain equal to or less than LRMC values. Given that the cost recovery occurs under the LRMC pricing method, and LRMC is always greater than or equal to SRMC, when SRMC is less than LRMC implies a failure in the cost recovery. This results in a missing money problem, indicating that investment costs can not be recovered under the SRMC pricing method. Consequently, decision-makers and investors must carefully consider the chosen pricing method to determine the likelihood of recovering their invested capital. 

Table \ref{tab:SRMC} shows the LRMC profiles, the associated SRMC profiles, and whether cost recovery occurs. We can see that in Profiles 1 and 4, where LRMC and SRMC are equal, cost recovery occurs. However, for the other profiles, it does not. For example, consider the third SRMC profile in Table \ref{tab:SRMC}. The revenue is given by:  
\begin{align}
    & \text{Revenue} = D_o \cdot CP_g + D_p \cdot CP_g 
\end{align}
The associated total cost is calculated according to Equation \ref{eq:Generation cost P3}. Accordingly, the profit is calculated as:
\begin{align}
    & \text{Profit} = D_o \cdot CP_g + D_p \cdot CP_g \\
    & - (D_o \cdot (CI_g+2CP_g)+(D_p-D_o) \cdot (CI_g+CP_g)) = -D_p \cdot CI_g < 0 \notag
\end{align}
The profit is negative, indicating that only the operational costs are recovered, while the investment costs are not. 

\begin{table}[htbp]
    \centering
    \caption{LRMC profiles, associated SRMC profiles, and feasibility of cost recovery}
    \label{tab:SRMC}
    \begin{adjustbox}{width=1\textwidth}
    \begin{tabular}{p{0.05\textwidth} >{\centering\arraybackslash}p{0.25\textwidth} >{\centering\arraybackslash}p{0.25\textwidth} >{\centering\arraybackslash}p{0.13\textwidth}|>{\centering\arraybackslash}p{0.10\textwidth} >{\centering\arraybackslash}p{0.10\textwidth} >{\centering\arraybackslash}p{0.13\textwidth}}
        \toprule 
         & \multicolumn{2}{>{\centering\arraybackslash}p{0.45\textwidth}}{LRMC profile} & Cost \hspace{7pt} recovery & \multicolumn{2}{>{\centering\arraybackslash}p{0.2\textwidth}}{SRMC profile} & Cost \hspace{7pt} recovery \\
        \midrule 
        1 & $CL$ & $CL$ & \checkmark & $CL$ & $CL$ & \checkmark \\
        2 & $CL$ & $CI_g + 2CP_g - CL$ & \checkmark & $CL$ & $CP_g$ & \\
        3 & $CP_g$ & $CI_g + CP_g$ & \checkmark & $CP_g$ & $CP_g$ & \\
        4 & $CL$ & $CP_g$ & \checkmark & $CL$ & $CP_g$ & \checkmark \\
        5 & $CL$ & $CI_g + CP_g$ & \checkmark & $CL$ & $CP_g$ & \\
        6 & $CP_r$ & $CI_f + CP_f$ & \checkmark & $CP_r$ & $CP_f$ & \\
        7 & $CI_f + 2CP_f - (CI_r + CP_r)$ & $CI_r + CP_r$ & \checkmark & $CP_f$ & $CP_r$ & \\
        \bottomrule 
    \end{tabular}
    \end{adjustbox}
\end{table}

\section{Concluding remarks} \label{sec:Concluding remarks}

In this paper, we have examined the costs of electricity generation and their alignment with optimal resource allocation, efficient pricing strategies, and cost recovery. This study investigates the relationship between long-run marginal cost (LRMC) and short-run marginal cost (SRMC), analyzes cost formation, and assesses cost recovery under each pricing method.

To address these issues, we focused on a simplified power system featuring two types of generators and two time periods. Subsequently, we devised a five-step methodology designed to estimate the LRMCs within this power system for different instance groups. Based on the results, we identified all potential LRMC combinations and explored the circumstances under which these combinations occur. 

Our analysis of the LRMC models demonstrates that generation cost recovery, including both investment and operational costs, is assured under the LRMC pricing method. In terms of investment cost distribution, if a generator is used solely within one period, its total investment cost is assigned to that period. However, if a generator is used in both peak and off-peak periods, the investment cost is distributed across these periods. In some cases, the entire investment cost is allocated to the peak period, particularly when the marginal generator's investment cost in the off-peak period is lower than or equal to the investment cost of the marginal generator in the peak period. If the investment cost of the marginal generator in the off-peak period is higher, an amount equal to the marginal generator's investment cost in the peak period is allocated to the peak period, and the remainder is allocated to the off-peak period. Similarly, during load shedding in the peak period, if the marginal generator's investment cost is less than or equal to the loadshed cost, it is fully allocated to the peak period. Otherwise, only a portion equal to the loadshed cost is allocated to the peak period, with the remainder assigned to the off-peak period. This method ensures efficient cost recovery and proper distribution of investment costs.

According to the results of the SRMC models, it is important to address degeneracy when calculating SRMC. Primal degeneracy in the SRMC model led to multiple SRMC values, none of which provided a reliable indication of the SRMC. To overcome this challenge, we resolved the degeneracy issue before calculating the SRMCs. The study highlights that SRMCs are consistently equal to or less than LRMCs, signaling a potential challenge in cost recovery. This underscores the importance of carefully selecting the pricing method to ensure cost recovery. Moreover, SRMC can take values equivalent to either the operational cost or the loadshed cost. In this pricing method, the focus is on immediate and variable expenses directly connected to electricity generation. This approach enables a flexible assessment of costs, allowing power generators to quickly adapt to short-term changes in demand or operational conditions.

Future research efforts should consider the intricate dynamics of the power system by incorporating storage and transmission. Examining their contributions to both LRMC and SRMC would contribute to a more realistic model, yielding more accurate prices. Additionally, introducing a more realistic assumption, such as price-responsive demand, can be of significance. This entails acknowledging that demand can dynamically shift in response to price fluctuations, impacting both LRMC and SRMC. Furthermore, expanding the scope to include carbon-emission caps adds a layer of environmental consideration to the model. Understanding the interplay between carbon emissions, costs, and pricing can provide valuable insights into the environmental sustainability of electricity generation practices. Lastly, exploring non-continuous investment capacities acknowledges the real-world constraints that may limit the expansion or modification of power generation facilities.

\section*{Acknowledgments}

This publication has been produced with support from the NCCS Research Centre, performed under the Norwegian research program Centres for Environment-friendly Energy Research (FME). The authors acknowledge the following partners for their contributions: Aker Carbon Capture, Ansaldo Energia, Baker Hughes, CoorsTek Membrane Sciences, Equinor, Elkem, Eramet, Fortum Oslo Varme, Gassco, KROHNE, Larvik Shipping, Lundin Norway, Norcem, Norwegian Oil and Gas, Quad Geometrics, Stratum Reservoir, Total Energies, Vår Energi, Wintershall DEA and the Research Council of Norway (257579).

\appendix

\appendix
\setcounter{table}{0} 
\renewcommand{\thetable}{A\arabic{table}} 
\renewcommand{\theequation}{A\arabic{equation}} 
\setcounter{equation}{0} 

\section*{Appendix}
\subsection*{Appndix A. Supplementary tables} \label{sec:supplementaryTables}
All instance groups and their characteristics are presented in Tables \ref{tab:identification} and \ref{tab:instance_groups}.

\begin{sidewaystable}[htbp]
    \centering
    \caption{Clusters, identification factors, and instance groups}
    \label{tab:identification}
    \begin{adjustbox}{width=0.9\textwidth}
    \begin{tabular}{c c c c c c c }
        \toprule
        \multirow{2}{*}{Cluster} & \multicolumn{4}{c }{Identification factors} & \multirow{2}{*}{Instance group} \\
        \cline{2-5}
        & Factor 1 & Factor 2 & Factor 3 & Factor 4 & \\
        \midrule
        1 & $D_1 < D_2$ & $D_1 < Mr$ & $CL < CI_r/2 + CP_r$ & - & 1 \\
        & & & $CI_r/2 + CP_r < CL < CI_r + CP_r$ & - & 2 \\
        & & & $CI_r + CP_r < CL < CI_f + CP_f$ & $D_1 < D_2 < D_1 + Mr$ & 3 \\
        & & & $CI_r + CP_r < CL < CI_f + CP_f$ & $D_1 + Mr < D_2 < 2Mr$ & 4 \\
        & & & $CI_r + CP_r < CL < CI_f + CP_f$ & $D_2 > 2Mr$ & 5 \\
        & & & $CL > CI_f + CP_f$ & $2Mr < D_2 < 2Mr + M_f$ & 6 \\
        & & & $CL > CI_f + CP_f$ & $2Mr + M_f < D_2 < 2Mr + 2M_f$ & 7 \\
        & & & $CL > CI_f + CP_f$ & $D_2 > 2Mr + 2M_f$ & 8 \\
        \midrule
        2 & $D_1 < D_2$ & $Mr < D_1 < Mr + M_f$ & $CL < CI_r/2 + CP_r$ & - & 9 \\
        & & & $CI_r/2 + CP_r < CL < CI_f/2 + CP_f$ & - & 10 \\
        & & & $CI_f/2 + CP_f < CL < CI_r + CP_r$ & - & 11 \\
        & & & $CI_r + CP_r < CL < CI_f + CP_f$ & $D_1 < D_2 < D_1 + Mr$ & 12 \\
        & & & $CI_r + CP_r < CL < CI_f + CP_f$ & $D_2 > D_1 + Mr$ & 13 \\
        & & & $CL > CI_f + CP_f$ & $D_1 + Mr < D_2 < D_1 + Mr + M_f$ & 14 \\
        & & & $CL > CI_f + CP_f$ & $D_1 + Mr + M_f < D_2 < D_1 + Mr + 2M_f$ & 15 \\
        & & & $CL > CI_f + CP_f$ & $D_2 > 2Mr + 2M_f$ & 16 \\
        \midrule
        3 & $D_1 < D_2$ & $D_1 > Mr + M_f$ & $CL < CI_r/2 + CP_r$ & - & 17 \\
        & & & $CI_r/2 + CP_r < CL < CI_f/2 + CP_f$ & - & 18 \\
        & & & $CI_f/2 + CP_f < CL < CI_r + CP_r$ & - & 19 \\
        & & & $CI_r + CP_r < CL < CI_f + CP_f$ & $Mr + M_f < D_2 < 2Mr + M_f$ & 20 \\
        & & & $CI_r + CP_r < CL < CI_f + CP_f$ & $D_2 > 2Mr + M_f$ & 21 \\
        & & & $CL > CI_f + CP_f$ & $2Mr + M_f < D_2 < 2Mr + 2M_f$ & 22 \\
        & & & $CL > CI_f + CP_f$ & $D_2 > 2Mr + 2M_f$ & 23 \\
        \midrule
        4 & $D_2 < D_1$ & $D_2 < Mr$ & $CL < CI_r/2 + CP_r$ & - & 24 \\
        & & & $CI_r/2 + CP_r < CL < CI_r + CP_r$ & - & 25 \\
        & & & $CI_r + CP_r < CL < CI_f + CP_f$ & $D_2 < D_1 < Mr$ & 26 \\
        & & & $CI_r + CP_r < CL < CI_f + CP_f$ & $D_1 > Mr$ & 27 \\
        & & & $CL > CI_f + CP_f$ & $Mr < D_1 < Mr + M_f$ & 28 \\
        & & & $CL > CI_f + CP_f$ & $D_1 > Mr + M_f$ & 29 \\
        \midrule
        5 & $D_2 < D_1$ & $Mr < D_2 < Mr + M_f$ & $CL < CI_r/2 + CP_r$ & - & 30 \\
        & & & $CI_r/2 + CP_r < CL < CI_f/2 + CP_f$ & - & 31 \\
        & & & $CI_f/2 + CP_f < CL < CI_f + CP_f$ & - & 32 \\
        & & & $CL > CI_f + CP_f$ & $D_2 < D_1 < Mr + M_f$ & 33 \\
        & & & $CL > CI_f + CP_f$ & $D_1 > Mr + M_f$ & 34 \\
        \midrule
        6 & $D_2 < D_1$ & $D_2 > Mr + M_f$ & $CL < CI_r/2 + CP_r$ & - & 35 \\
        & & & $CI_r/2 + CP_r < CL < CI_f/2 + CP_f$ & - & 36 \\
        & & & $CI_f/2 + CP_f < CL < CI_f + CP_f$ & - & 37 \\
        & & & $CI_r + CP_r < CL < CI_f + CP_f$ & $Mr + M_f < D_2 < 2Mr + M_f$ & 38 \\
        & & & $CI_r + CP_r < CL < CI_f + CP_f$ & $D_2 > 2Mr + M_f$ & 39 \\
        & & & $CL > CI_f + CP_f$ & $2Mr + M_f < D_2 < 2Mr + 2M_f$ & 40 \\
        & & & $CL > CI_f + CP_f$ & $D_2 > 2Mr + 2M_f$ & 41 \\
        \bottomrule
    \end{tabular}
    \end{adjustbox}
\end{sidewaystable}

\begin{sidewaystable}
    \centering
    \caption{Instance groups and their characteristics}
    \label{tab:instance_groups}
    \begin{adjustbox}{width=\textwidth}
    \begin{tabular}{ccccccccccccc}
        \toprule
        Ins. gr. & \multicolumn{2}{c}{Used generation option} & $I_{r1}$ & $I_{r2}$ & $I_{f1}$ & $I_{f2}$ & $L_1$ & $L_2$ & $\lambda_1$ & $\lambda_2$ & LRMC profile \\
        \cline{2-3}
        & Period 1 & Period 2 & & & & & & & & &\\
        \midrule
        1 & - & - & 0 & 0 & 0 & 0 & $D_1$ & $D_2$ & $CL$ & $CL$ & Profile 1\\
        2 & SR & SR & $D_1$ & 0 & 0 & 0 & 0 & $D_2-D_1$ & $CI_r+2CP_r-CL$ & $CL$ & Profile 2 \\
        3 & SR & SR, R & $D_1$ & $D_2-D_1$ & 0 & 0 & 0 & 0 & $CP_r$ & $CI_r+CP_r$ & Profile 3 \\
        4 & SR, R & SR, R & $D_2-M_r$ & $M_r$ & 0 & 0 & 0 & 0 & $CP_r$ & $CI_r+CP_r$ & Profile 3 \\
        5 & SR, R & SR, R & $M_r$ & $M_r$ & 0 & 0 & 0 & $D_2-2M_r$ & $CP_r$ & $CL$ & Profile 4 \\
        6 & SR, R & SR, R, F & $M_r$ & $M_r$ & 0 & $D_2-2M_r$ & 0 & 0 & $CP_r$ & $CI_f+CP_f$ & Profile 6 \\
        7 & SR, R, F & SR, R, F & $M_r$ & $M_r$ & $D_2-2M_r-M_f$ & $M_f$ & 0 & 0 & $CP_r$ & $CI_f+CP_f$ & Profile 6 \\
        8 & SR, R, F & SR, R, F & $M_r$ & $M_r$ & $M_f$ & $M_f$ & 0 & $D_2-2M_r-2M_f$ & $CP_r$ & $CL$ & Profile 4 \\
        \midrule
        9 & - & - & 0 & 0 & 0 & 0 & $D_1$ & $D_2$ & $CL$ & $CL$ & Profile 1 \\
        10 & SR & SR & $M_r$ & 0 & 0 & 0 & $D_1-M_r$ & $D_2-M_r$ & $CL$ & $CL$ & Profile 1 \\
        11 & SR, SF & SR, SF & $M_r$ & 0 & $D_1-M_r$ & 0 & 0 & $D_2-D_1$ & $CI_f+2CP_f-CL$ & $CL$ & Profile 2 \\
        12 & SR, SF & SR, SF, R & $M_r$ & $D_2-D_1$ & $D_1-M_r$ & 0 & 0 & 0 & $CI_f+2CP_f-(CI_r+CP_r)$ & $CI_r+CP_r$ & Profile 7 \\
        13 & SR, SF & SR, SF, R & $M_r$ & $M_r$ & $D_1-M_r$ & 0 & 0 & $D_2-(D_1-M_r)$ & $CI_f+2CP_f-CL$ & $CL$ & Profile 2 \\
        14 & SR, SF & SR, SF, R, F & $M_r$ & $M_r$ & $D_1-M_r$ & $D_2-(D_1-M_r)$ & 0 & 0 & $CP_f$ & $CI_f+CP_f$ & Profile 3 \\
        15 & SR, SF, F & SR, SF, R, F & $M_r$ & $M_r$ & $D_2-2M_r-M_f$ & $M_f$ & 0 & 0 & $CP_f$ & $CI_f+CP_f$ & Profile 3 \\
        16 & SR, SF, F & SR, SF, R, F & $M_r$ & $M_r$ & $M_f$ & $M_f$ & 0 & $D_2-2M_r-2M_f$ & $CP_f$ & $CL$ & Profile 4 \\
        \midrule
        17 & - & - & 0 & 0 & 0 & 0 & $D_1$ & $D_2$ & $CL$ & $CL$ & Profile 1 \\
        18 & SR & SR & $M_r$ & 0 & 0 & 0 & $D_1-M_r$ & $D_2-M_r$ & $CL$ & $CL$ & Profile 1 \\
        19 & SR, SF & SR, SF & $M_r$ & 0 & $M_f$ & 0 & $D_1-M_r-M_f$ & $D_2-M_r-M_f$ & $CL$ & $CL$ & Profile 1 \\
        20 & SR, SF & SR, SF, R & $M_r$ & $D_2-M_r-M_f$ & $M_f$ & 0 & $D_1-M_r-M_f$ & 0 & $CL$ & $CI_r+CP_r$ & Profile 5 \\
        21 & SR, SF & SR, SF, R & $M_r$ & $M_r$ & $M_f$ & 0 & $D_1-M_r-M_f$ & $D_2-2M_r-M_f$ & $CL$ & $CL$ & Profile 1 \\
        22 & SR, SF & SR, SF, R, F & $M_r$ & $M_r$ & $M_f$ & $D_2-2M_r-M_f$ & $D_1-M_r-M_f$ & 0 & $CL$ & $CI_f+CP_f$ & Profile 5 \\
        23 & SR, SF & SR, SF, R, F & $M_r$ & $M_r$ & $M_f$ & $M_f$ & $D_1-M_r-M_f$ & $D_2-2M_r-2M_f$ & $CL$ & $CL$ & Profile 1 \\
        \midrule
        24 & - & - & 0 & 0 & 0 & 0 & $D_1$ & $D_2$ & $CL$ & $CL$ & Profile 1 \\
        25 & SR & SR & $D_2$ & 0 & 0 & 0 & $D_1-D_2$ & 0 & $CL$ & $CI_r+2CP_r-CL$ & Profile 2 \\
        26 & SR, R & SR & $D_1$ & 0 & 0 & 0 & 0 & 0 & $CI_r+CP_r$ & $CP_r$ & Profile 3 \\
        27 & SR, R & SR & $M_r$ & 0 & 0 & 0 & $D_1-M_r$ & 0 & $CL$ & $CP_r$ & Profile 4 \\
        28 & SR, R, F & SR & $M_r$ & 0 & $D_1-M_r$ & 0 & 0 & 0 & $CI_f+CP_f$ & $CP_r$ & Profile 6 \\
        29 & SR, R, F & SR & $M_r$ & 0 & $M_f$ & 0 & $D_1-M_r-M_f$ & 0 & $CL$ & $CP_r$ & Profile 4 \\
        \midrule
        30 & - & - & 0 & 0 & 0 & 0 & $D_1$ & $D_2$ & $CL$ & $CL$ & Profile 1 \\
        31 & SR & SR & $M_r$ & 0 & 0 & 0 & $D_1-M_r$ & $D_2-M_r$ & $CL$ & $CL$ & Profile 1 \\
        32 & SR, SF & SR & $M_r$ & 0 & $D_2-M_r$ & 0 & $D_1-M_r$ & 0 & $CL$ & $CI_r+2CP_r-CL$ & Profile 2 \\
        33 & SR, SF, F & SR & $M_r$ & 0 & $D_1-M_r$ & 0 & 0 & 0 & $CI_f+CP_f$ & $CP_f$ & Profile 3 \\
        34 & SR, SF, F & SR & $M_r$ & 0 & $M_f$ & 0 & $D_1-M_r-M_f$ & 0 & $CL$ & $CP_f$ & Profile 4 \\
        \midrule
        35 & - & - & 0 & 0 & 0 & 0 & $D_1$ & $D_2$ & $CL$ & $CL$ & Profile 1 \\
        36 & SR & SR & $M_r$ & 0 & 0 & 0 & $D_1-M_r$ & $D_2-M_r$ & $CL$ & $CL$ & Profile 1 \\
        37 & SR, SF & SR, SF & $M_r$ & 0 & $M_f$ & 0 & $D_1-M_r-M_f$ & $D_2-M_r-M_f$ & $CL$ & $CL$ & Profile 1 \\
        38 & SR, SF & SR, SF, R & $M_r$ & $D_2-M_r-M_f$ & $M_f$ & 0 & $D_1-M_r-M_f$ & 0 & $CL$ & $CI_r+CP_r$ & Profile 5 \\
        39 & SR, SF & SR, SF, R & $M_r$ & $M_r$ & $M_f$ & 0 & $D_1-M_r-M_f$ & $D_2-2M_r-M_f$ & $CL$ & $CL$ & Profile 1 \\
        40 & SR, SF & SR, SF, R, F & $M_r$ & $M_r$ & $M_f$ & $D_2-2M_r-M_f$ & $D_1-M_r-M_f$ & 0 & $CL$ & $CI_f+CP_f$ & Profile 5 \\
        41 & SR, SF & SR, SF, R, F & $M_r$ & $M_r$ & $M_f$ & $M_f$ & $D_1-M_r-M_f$ & $D_2-2M_r-2M_f$ & $CL$ & $CL$ & Profile 1 \\
        \bottomrule
    \end{tabular}
    \end{adjustbox}
\end{sidewaystable}

\subsection*{Appendix B. Degeneracy} \label{sec:degeneracy}
If we want to be more clear about why degeneracy happens, we can consider the objective function of the dual model for SRMC as follows. We suppose that the first period is the off-peak period. 
\begin{align}
    \max y &= CI_g \cdot I^*_{g1} + CI_g \cdot I^*_{g2} + D_1 \cdot \dot{\lambda_{1}} + D_2 \cdot \dot{\lambda_{2}} - I^*_{g1} \cdot \dot{\beta_{g1}} - (I^*_{g1} + I^*_{g2}) \cdot \dot{\beta_{g2}} &&
\end{align}
If we substitute $\dot{\beta_{g1}}$ in the objective function ($\dot{\beta_{g1}}= \dot{\lambda_1} - CP_g$), the coefficient of $\dot{\lambda_1} $ becomes $D_1 - I^*_{g1}$. It is known that the demand in the first period is lower than that in the second period. The investment made in the initial period equates to the capacity invested in generator $g$ for that period. This investment amount corresponds to the shared capacity of generator $g$, which is given by $\min \{D_1, D_2\} = D_1$. Consequently, $I^*_{g1}=D_1$, resulting in a zero coefficient for $\dot{\lambda_1}$ within the objective function. Thus, $\dot{\lambda_1}$ can take all values between $CP_g$ and $CL$, signifying the primal degeneracy and presence of multiple solutions within the dual model.

To resolve the degeneracy and derive an accurate value for the SRMC, we add a small positive value denoted by $\epsilon$ to the fixed investment capacities, yielding an optimal value of $I^* + \epsilon$ as the adjusted investment capacities. In fact, SRMC is equivalent to the additional cost associated with the generation of an extra unit of electricity when ample investment capacity exists. By adding $\epsilon$ value, we adjust the model to incorporate the notion of sufficient capacity and subsequently proceed to compute the SRMC. After the adjustment, the complementary slackness conditions will be written as follows:  
\begin{align}
&P_{g1} \cdot (\dot{\lambda_{1}} - \dot{\beta_{g1}} - CP_g) = 0 : &&P_{g1}>0 && \Rightarrow \dot{\lambda_1} - \dot{\beta_{g1}} = CP_g \\
&\dot{\beta_{g1}} \cdot (-P_{g1} + I^*_{g1}+\epsilon) = 0 : && P_{g1}<I^*_{g1}+\epsilon && \Rightarrow \dot{\beta_{g1}} = 0 \notag
\end{align}
Consequently, through the resolution of degeneracy, it is concluded that $\dot{\lambda_{1}} = CP_g$.

\bibliographystyle{elsarticle-harv} 
\bibliography{elsarticle/references}



\end{document}